\RequirePackage{fix-cm}
\documentclass{svjour3}                     % onecolumn (standard format)
\smartqed  % flush right qed marks, e.g. at end of proof
\usepackage{graphicx}

\usepackage{amsmath,amsthm,amsfonts,amssymb,bbm,bm,mathtools}
\usepackage{caption}
\usepackage{subcaption}
\usepackage{float}
\usepackage{tikz}
\usepackage{tikz-network}
\usetikzlibrary{arrows}
\tikzset{>=latex}
\usetikzlibrary{patterns,decorations.pathreplacing}
\usepackage{color}
\definecolor{myred}{RGB}{232,102,102}
\definecolor{myblue}{RGB}{187,187,255}
\definecolor{mygreen}{RGB}{34,139,34}
\definecolor{myorange}{RGB}{255,165,0}
\definecolor{OliveGreen}{RGB}{85,107,47}
\definecolor{NavyBlue}{RGB}{0,0,128}
\usepackage[colorlinks,bookmarks=false,citecolor=NavyBlue,linkcolor=OliveGreen,urlcolor=blue]{hyperref}
\usepackage{tensor}

\newcommand{\be}{\begin{equation}}
\newcommand{\ee}{\end{equation}}
\newcommand{\bea}{\begin{eqnarray}}
\newcommand{\eea}{\end{eqnarray}}

\newcommand{\one}{\mathbbm{1}}
\newcommand{\ave}[1]{{\langle #1 \rangle}}

\journalname{Journal of Statistical Physics}
\begin{document}

\title{Kardar-Parisi-Zhang physics in integrable rotationally symmetric dynamics on discrete space-time lattice%\thanks{Grants or other notes
}
\subtitle{}
\titlerunning{KPZ physics in integrable SO(3) symmetric dynamics on discrete space-time lattice}


\author{\v Ziga Krajnik         \and
        Toma\v z Prosen %etc.
}


\institute{
              Physics Department, Faculty of Mathematics and Physics, University of Ljubljana, Jadranska 19, SI-1000,
Ljubljana, Slovenia \\
              \email{tomaz.prosen@fmf.uni-lj.si}           
}

\date{Preprint, dated: \today}

\maketitle

\begin{abstract}
We introduce a deterministic SO(3) invariant dynamics of classical spins on a discrete space-time lattice and prove its complete integrability by explicitly finding a related 
non-constant (baxterized) solution of the set-theoretic quantum Yang-Baxter equation over the 2-sphere. 
Equipping the algebraic structure with the corresponding Lax operator we derive an infinite sequence of conserved quantities with local densities.
The dynamics depend on a single continuous spectral parameter and reduce to a (lattice) Landau-Lifshitz model in the limit of a small parameter which corresponds to the continuous time limit.
Using quasi-exact numerical simulations of deterministic dynamics and Monte Carlo sampling of initial conditions corresponding to a maximum entropy equilibrium state
we determine spin-spin spatio-temporal (dynamical) correlation functions with relative accuracy of three orders of magnitude. 
We demonstrate that in the equilibrium state with a vanishing total magnetization the correlation function precisely follow Kardar-Parisi-Zhang scaling hence the spin transport belongs to the
universality class with dynamical exponent $z=3/2$, in accordance to recent related simulations in discrete and continuous time quantum Heisenberg spin 1/2 chains.
\end{abstract}

\section{Introduction}
\label{intro}
Identifying exactly solvable cases of universal physical phenomena is one of the central goals of statistical physics. While this endeavour has matured in equilibrium physics it is still at its very early stage for non-equilibrium phenomena such as transport. 
Considering interacting particle models without any hidden degrees of freedom, meaning that the underlying microscopic equations of motion are deterministic and reversible\footnote{Excluding stochastic systems or any external sources of noise.}, establishing macroscopic transport laws, such as Fick's or Ohm's law is particularly hard, while only very 
recently fully explicit and rigorous results started to emerge (see e.g. \cite{Katja}). On a more heuristic level, distinct types of transport phenomena, related e.g. to two important universality classes given by the 
diffusion equation and the Kardar-Parisi-Zhang \cite{KPZ} (KPZ) equation, have been explained via {\em nonlinear fluctuating hydrodynamics} (NFH) which crucially depends on the number of conserved fields 
(such as mass, energy or momentum densities) and nonlinear coupling relations among their currents. 
With this heuristic theory, one can predict either diffusive or anomalous broadening  of the (moving) sound-peaks and (static) sound-peak in the 
space-time resolved dynamical response functions and their precise asymptotic scaling profiles \cite{Spohn,Spohn2,Popkov}. 
Whenever {\rm three appropriately coupled} conserved fields have been identified, such as in the Femi-Pasta-Ulam problem \cite{MendlSpohn}, the mean-field description of Bose gasses \cite{Lamacraft}, 
or classical XXZ spin chains at low temperatures \cite{DharSpohn1}, the broadenings have been explained in terms of the KPZ scaling $x \sim t^{1/z}$ with dynamical exponent $z=3/2$, in distinction to diffusive 
and ballistic universality classes with exponents $z=2$ and $z=1$ which, respectively, typically emerge in `more generic' chaotic or integrable models. 

Very recently, however, a new kind of incarnation of KPZ physics in deterministic statistical systems has been suggested. Specifically, studying dynamical correlations in quantum Heisenberg (XXX) chain of spins 1/2 
at vanishing magnetization (or zero magnetic field), it has been demonstrated that the sole contribution to transport comes from the heat-peak with the sound-peaks being absent, but that the former 
broadens with a perfect KPZ scaling \cite{quant-KPZ} over several orders of magnitude. This result is consistent with earlier observations \cite{LLL_numerics}  of $z\sim1.5$ in the integrable lattice Landau-Lifshitz (LLL) chain of classical spins 
\cite{Faddeev-Takhtajan} which can be thought of as an integrable classical version of the XXX model. More recently, these numerical experiments have been refined, confirming also the precise KPZ scaling profile of the heat-peak \cite{DharSpohn2}. 
Such a behaviour seem to crucially depend on the complete integrability and on rotational (SO(3)) symmetry, but not on its quantum or classical nature. 
While hydrodynamics has recently been generalized to integrable systems with infinite number of conservation laws \cite{GHD1,GHD2} (so-called GHD), where diffusion and anomalous diffusion could be included with some heuristics \cite{Jacopo,Romain}, 
it is not clear at present how and if one can explain the observed behavior within NFH. From empirical observations, it seems that for this type of KPZ scaling one needs two ingredients: (i) complete integrability and (ii) existence of a global non-abelian symmetry (such as SU(2)) with non-commuting conserved generators. See Ref.~\cite{Joel} for additional data corroborating this conjecture for higher spin integrable quantum models with higher rank nonabelian symmetries.
Since low energy regimes of non-integrable lattice models are often described by integrable field theories, one could then apply such results also to non-integrable SO(3) symmetric lattice models at low temperatures \cite{Enej2} or integrable SO(3) symmetric field theories \cite{Enej}.

In order to prepare the stage for a rigorous case study, as well as to sharpen numerical evidence as much as possible, we are here defining and studying arguably the simplest dynamical system satisfying the required conditions and demonstrate 
that it indeed exhibits KPZ scaling.

In the first part of the paper (section 2) we introduce a discrete time dynamical system of $N$ normalized angular momenta (classical spins), each taking values on a 2-sphere ${\cal S}^2$, 
which is generated by a simple symplectic (canonical) transformation over $({\cal S}^2)^{\times N}$. This many-body map has the form of a classical Floquet circuit built from a simple local 2-spin mapping which is a simple non-linear rational and rotationally symmetric
bijective transformation of a pair of unit 3-vectors. Such a dynamical system should be of interest in its own right, since we demonstrate that the 2-spin mapping satisfies a baxterized set theoretic {\em quantum} Yang-Baxter equation where the spectral parameter plays the role of the `integration time-step'. Moreover, we introduce an appropriate Lax matrix and prove the corresponding set theoretic quantum RLL relation which allows us to compute an extensive family of conserved fields of the model.
The model therefore represents the simplest known rotationally (SO(3)) symmetric integrable dynamics in discrete-space time and due to its efficient simulability provides a perfect playground for testing the above phenomenological conjecture on the KPZ scaling.
Moreover, we show that our dynamics exhibits a remarkable space-time symmetry, namely it is generated by essentially the same deterministic and reversible many-body map if one flips the time and space axes. In other words, knowing the value of a fixed spins at all moments in time, we can find (via `space dynamics') unique values of all other spins at all time steps. 

In the second part of the paper (section 3) we then numerically explore dynamical spin-spin correlation functions in the simplest separable invariant state (which can be understood as an infinite temperature/maximum entropy state at fixed average magnetization) and demonstrate that it obeys a clean KPZ scaling for vanishing magnetization. When changing the magnetization parameter we then demonstrate a crossover to a ballistic scaling which could be captured within GHD.
Moreover, when slightly breaking integrability of the discrete-time mapping while keeping the same continuous time limit (namely, the LLL model), we demonstrate an immediate drift of dynamical exponents towards the diffusive value $z=2$.

\section{Integrable ${\rm SO}(3)$ invariant dynamics on a discrete space-time lattice}
\subsection{Definition of the model}
\label{sec:krajevni1-definition}
\begin{figure}
\centering
\begin{tikzpicture}[baseline=(current  bounding  box.center), scale=1.2]
% Spin labels
\foreach \i in {2}
{
\Text[x=-11.7,y=\i -4.3]{\large$\mathbf{S}_{2x}^{2t+\i}$}
}

\foreach \i in {0}
{
\Text[x=-11.8,y=\i -4.3]{\large$\mathbf{S}_{2x}^{2t}$}
}

\foreach \i in {1, 3}
{
\Text[x=-11.7,y=\i -3.9]{\large$\mathbf{S}_{2x}^{2t+\i}$}
}
\foreach \i in {2, 4}
{
\Text[x=-11.35 + \i,y=-4.3]{\large$\mathbf{S}_{2x+\i}^{2t}$}
}
\foreach \i in {1, 3, 5}
{
\Text[x=-11.2 + \i,y=-4.3]{\large$\mathbf{S}_{2x+\i}^{2t}$}
}

% Triple dots
\Text[x=-5.5, y=-4.2]{$\boldsymbol{\cdots}$}
\Text[x=-11.8, y=0.0]{$\boldsymbol{\vdots}$}

% Coordinate axes
\draw[thick, black, ->] (-5.2,-4.0) -- node [left, pos=.2]  {\large$t$} (-5.2,-1.8);
\draw[thick, black, ->] (-11.5, 0.2) -- node [below, pos=.2] {\large$x$}  (-9.1,0.2);

% Leftover black dots
\draw[thick, fill=black] (-11.5, -1) circle (0.05cm); 
\draw[thick, fill=black] (-5.5, -2) circle (0.05cm); 
\draw[thick, fill=black] (-5.5, -3) circle (0.05cm); 

\foreach \i in {3,...,5}
{
\draw[thick] (-.5-2*\i,-1) -- (0.525-2*\i,0.025);
\draw[thick] (-0.525-2*\i,0.025) -- (0.5-2*\i,-1);
\draw[thick, fill=black] (-16.5 + 2*\i,-1) circle (0.05cm); 
\draw[thick, fill=black] (-15.5 + 2*\i,-1) circle (0.05cm); 
\draw[thick, fill=black] (-16.5 + 2*\i,0) circle (0.05cm); 
\draw[thick, fill=black] (-15.5 + 2*\i,0) circle (0.05cm); 
\draw[thick, fill=white, rounded corners=2pt] (-0.25-2*\i,-0.25) rectangle (.25-2*\i,-0.75) node[pos=.5]{\Large $\, \Phi_{\tau}$};
\draw[thick, fill=mygreen, fill opacity=0.4, rounded corners=2pt] (-0.25-2*\i,-0.25) rectangle (.25-2*\i,-0.75);
}

\foreach \jj[evaluate=\jj as \j using -2*(ceil(\jj/2)-\jj/2)] in {0,...,2}
\foreach \i in {4,...,6}
{
\draw[thick, fill=black] (-19.5 + 2*\i,-2-\jj) circle (0.05cm); 
\draw[thick, fill=black] (-18.5 + 2*\i,-2-\jj) circle (0.05cm); 
\draw[thick] (.5-2*\i-1*\j,-2-1*\jj) -- (1-2*\i-1*\j,-1.5-\jj);
\draw[thick] (1-2*\i-1*\j,-1.5-1*\jj) -- (1.5-2*\i-1*\j,-2-\jj);
}
\foreach \jj[evaluate=\jj as \j using -2*(ceil(\jj/2)-\jj/2)] in {0,...,2}
\foreach \i in {4,...,6}
{
\draw[thick] (.5-2*\i-1*\j,-1-1*\jj) -- (1-2*\i-1*\j,-1.5-\jj);
\draw[thick] (1-2*\i-1*\j,-1.5-1*\jj) -- (1.5-2*\i-1*\j,-1-\jj);
\draw[thick, fill=white, rounded corners=2pt] (0.75-2*\i-1*\j,-1.75-\jj) rectangle (1.25-2*\i-1*\j,-1.25-\jj) node[pos=.5]{\Large $\, \Phi_{\tau}$};
\draw[thick, fill=mygreen, fill opacity=0.4, rounded corners=2pt] (0.75-2*\i-1*\j,-1.75-\jj) rectangle (1.25-2*\i-1*\j,-1.25-\jj);
}

\end{tikzpicture}
 \caption{Classical local symplectic circuit representation of discrete space-time dynamics of the model.
 Lattice spins (taking values on 2-spheres) are denoted by black circles. A green rectangle represents a two-body propagator (\ref{two-body-map})  applied to a pair of spins. Time increases along the vertical axis, the spatial lattice runs along the horizontal axis.}
\label{fig:trotter}
\end{figure}
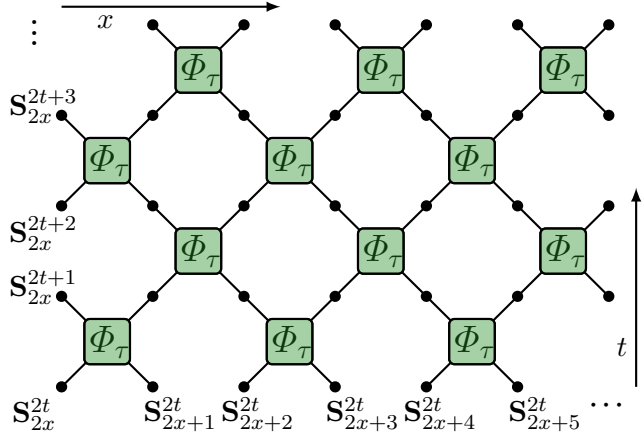

Let $\mathbf{S}_1, \mathbf{S}_2$ denote a pair of three-dimensional unit vectors, $\vec{S}_1\cdot\vec{S}_1=\vec{S}_2\cdot\vec{S}_2=1$.
We define a one parameter family of rational nonlinear maps $\Phi_{\tau}$ between a pair of 2-spheres 
$
\Phi_{\tau}  : S^2 \times S^2 \rightarrow  S^2 \times S^2, 
$
as:
\bea
\label{two-body-map}
\Phi_{\tau}  (\mathbf{S}_1, \mathbf{S}_2) &=&  \frac{1}{\sigma^2 + \tau^2} \Big(\sigma^2\mathbf{S}_1 + \tau^2\mathbf{S}_2 + \tau \mathbf{S}_1\times\mathbf{S}_2, \sigma^2\mathbf{S}_2 + \tau^2\mathbf{S}_1 + \tau \mathbf{S}_2\times\mathbf{S}_1 \Big),\quad\\
\sigma^2 &:=& \frac{1}{2} \Big(1 + \mathbf{S}_1\cdot \mathbf{S}_2 \Big), \nonumber
\eea
where $\tau\in\mathbb R$ is a real parameter, which will later be interpreted as the discretization time step.  
A simple calculation shows that the map $\Phi_\tau$ preserves the unit norm of the pair of vectors and is invertible, thus it represents a bijection  on 
${\cal S}^2\times {\cal S}^2$ which is clearly invariant under
rotations:
\be
(\mathbf{S}'_1,\mathbf{S}'_2)=\Phi_\tau(\mathbf{S}_1,\mathbf{S}_2) \;\;\Leftrightarrow\;\;
({R}\mathbf{S}'_1,{R}\mathbf{S}'_2)=\Phi_\tau({R}\mathbf{S}_1,{R}\mathbf{S}_2),
\qquad R\in {\rm SO}(3).
\ee
Eq.~(\ref{two-body-map}) defines the elementary two-body propagator of our model. Let us now proceed to a definition of a discrete-time dynamics for a lattice (chain) of an {\em even} number $N \in 2\mathbb N$ of unit vectors:
\be 
\vec{S}_x^t \in S^2, \quad x\in\mathbb Z_{N}, \quad t\in\mathbb Z,
\ee
which we define as follows:
\be
(\vec{S}^{2t+1}_{2x},\vec{S}^{2t+1}_{2x+1}) = \Phi_\tau(\vec{S}^{2t}_{2x},\vec{S}^{2t}_{2x+1}), \qquad
(\vec{S}^{2t+2}_{2x-1},\vec{S}^{2t+1}_{2x}) = \Phi_\tau(\vec{S}^{2t+1}_{2x-1},\vec{S}^{2t+1}_{2x}),
\label{st}
\ee
for integer space-time indices $x\in\mathbb Z_{N/2}$, $t\in\mathbb Z$ (see a schematic depiction in Fig.~\ref{fig:trotter}).

This prescription can be understood as a discrete-time, deterministic, reversible dynamical system generated by an invertible dynamical map $\Psi_\tau : {\cal M}\to{\cal M}$ over a product of $N$ 2-spheres, ${\cal M} = ({\cal S}^2)^{\times N}$, 
which is defined as a composition of an even and odd half-time step propagators:
\bea
(\vec{S}_0^{2t+2},\vec{S}_{1}^{2t+2},\dots,\vec{S}_{N-1}^{2t+2}) &=& {\Psi}_\tau(\vec{S}_0^{2t},\vec{S}_{1}^{2t},\dots,\vec{S}_{N-1}^{2t}), \label{eq:Psi} \label{MBmap}\\
{\Psi}_\tau &=& {\Psi}^{\rm odd}_\tau \circ {\Psi}^{\rm even}_\tau,\nonumber \\
{\Psi}^{\rm even}_\tau &=& \Phi_\tau^{\otimes N/2},\nonumber \\
{\Psi}^{\rm odd}_\tau &=& \eta^{-1} \circ {\Psi}^{\rm even}_\tau \circ \eta. \nonumber
\eea 
The map:
\be
{\eta}(\vec{S}_0,\vec{S}_{1},\ldots,\vec{S}_{N-2},\vec{S}_{N-1}) = (\vec{S}_{1},\vec{S}_2,\ldots,\vec{S}_{N-1},\vec{S}_0)
\ee 
is a periodic translation on a classical spin-ring ${\cal M}$. 
The tensor product of maps over a cartesian product of their domain sets is defined as
$(\Omega\otimes\Lambda)(\vec{x},\vec{y}) \equiv (\Omega(\vec{x}),\Lambda(\vec{y}))$. 

Note that this discrete space-time dynamics is a classical analog of a local quantum circuit representation of a Trotter decomposition of unitary Hamiltonian dynamics. Particularly, since as we will show below, $\Phi_\tau$ can be
generated by a suitable 2-spin Hamiltonian and hence the many-body map $\Psi$ is a canonical transformation which is generated by a suitable (periodically) time-dependent Hamiltonian. The model can thus also be interpreted as a classical local Floquet circuit.

\subsection{The Hamiltonian structure of the elementary two-body interaction}
Before demonstrating the integrability of the model we make a brief detour and show the Hamiltonian and symplectic character of the building blocks of the model.
We seek a Hamiltonian $H(\vec{S}_1,\vec{S}_2)$ that will generate the canonical transformation  ({\ref{two-body-map}}) through the equations of motion after a time specified by $\tau$.
Since the two-body interaction is invariant under ${\rm SO}(3)$, the Hamiltonian must be a function of the scalar invariant of a pair of unit vectors, i.e. it should be of the form:
\begin{equation}
H(\vec{S}_1,\vec{S}_2) =  2h(\sigma), \quad \sigma^2=\frac{1}{2}(1 + \mathbf{S}_1 \cdot \mathbf{S}_2).
\end{equation}
Since we aim to interpret vectors $\vec{S}_n$, $n\in\{1,2\}$, as classical angular momenta (which we shall simply refer to as `spins') we invoke the ${\rm SO}(3)$ Poisson bracket with the canonical relations:
\begin{equation}
\{S_{n;a}, S_{m;b} \} = \delta_{n,m} \sum_c \varepsilon_{abc} S_{n;c},
\end{equation}
where $\varepsilon_{abc}$ is the Levi-Civita symbol and $S_{n;a}$, $a\in\{1,2,3\}$ denote the three components of the vector $\vec{S}_n$.
Hamilton's equations of motion then take the form:
\begin{align}
\dot{\mathbf{S}}_n := \frac{{\rm d}\vec{S}_n}{{\rm d}t} = \{\mathbf{S}_n, H\},
\end{align}
where the Poisson bracket acts on every component of the vector $\mathbf{S}_j$. Explicitly, we have:
 \be
\label{2b_eom}
\dot{\mathbf{S}}_1 =  (\mathbf{S}_1 \times \mathbf{S}_2)\frac{h'(\sigma)}{2\sigma}\,,\quad
 \dot{\mathbf{S}}_2 = (\mathbf{S}_2 \times \mathbf{S}_1)\frac{h'(\sigma)}{2\sigma} = -\dot{\vec{S}}_1\,.
\ee
The pair of Eqs.~(\ref{2b_eom}) implies that the sum of the spins and their dot product is conserved in time.  
The equations of motion can be rewritten in a form such that the second vector in the cross product is of unit length $\vec{\Sigma} = (\vec{S}_1+\vec{S}_2)/(2\sigma)$:
 \be
\dot{\mathbf{S}}_n =   h'(\sigma) \mathbf{S}_n \times \vec{\Sigma} 
\ee
from which it is clearly seen that the spins rotate around their conserved sum with angular velocity $h'(\sigma):={\rm d}h(\sigma)/{\rm d}\sigma$, 
hence their time evolution can be explicitly expressed via Rodrigues' rotation formula:
\begin{align}
\label{explicit_rotation}
\mathbf{S}_1(t) &= \frac{1}{2}\Big[\mathbf{S}_1(0) (1 + \cos h' t) + \mathbf{S}_2(0)  (1- \cos h' t) + \mathbf{S}_1(0)  \times \mathbf{S}_2(0) \frac{\sin h' t}{\sigma}  \Big],\\
\mathbf{S}_2(t) &= \frac{1}{2}\Big[\mathbf{S}_2(0)  (1 + \cos h' t) + \mathbf{S}_1(0)  (1- \cos h' t) + \mathbf{S}_2(0)  \times \mathbf{S}_1(0) \frac{\sin h' t}{\sigma} \Big]. \nonumber
\end{align}
Comparing the pair of Eqs.~(\ref{explicit_rotation}) at time $t=1$ with Eq.~(\ref{two-body-map}) we conclude that the angular velocity must satisfy the following differential equation:
\begin{equation}
\tan h'(\sigma) = \frac{2\sigma\tau}{\tau^2-\sigma^2},
\end{equation}
with an explicit solution, unique up to the choice of the branch of the inverse tangent:
\be
h(\sigma) = \sigma \arctan\left( \frac{2\sigma\tau}{\tau^2-\sigma^2}\right) - \tau \log\left(\tau^2+\sigma^2\right).
\label{eq:localh}
\ee
This concludes the proof that $\Phi_\tau$ (\ref{two-body-map}) is a non-linear canonical  (aka symplectic) transformation generated in unit time by $h(\sigma)$.
Since the mapping along hamiltonian trajectory (\ref{explicit_rotation}) for $t=-1$, corresponding to (\ref{two-body-map}) with a flipped sign of $\tau$, has to generate the inverted
symplectic transformation, we have:
\be
(\Phi_\tau)^{-1} = \Phi_{-\tau}.
\ee

The full many-body map $\Psi$ (\ref{eq:Psi}) can thus be generated by a two-step classical Floquet protocol with a periodic time-dependent Hamiltonian:
\bea
H(t) &=& \begin{cases}
H_{\rm even},\; \;\lfloor t\rfloor \; {\rm even};\\
H_{\rm odd},\; \; \;\lfloor t\rfloor \; {\rm odd};
\end{cases}\quad t\in\mathbb R,
\label{eq:two-step}\\
H_{\rm even} &=& \sum_{x=0}^{N/2-1}  2h(\sigma_{2x,2x+1}),\qquad H_{\rm odd} = \sum_{x=1}^{N/2} 2h(\sigma_{2x-1,2x}), \\
\sigma_{x,x+1} &=& \sqrt{\frac{1}{2}(1+\vec{S}_{x}\cdot\vec{S}_{x+1})}.
\eea

\subsection{The limit of $\tau \rightarrow 0$}
\label{sec:1-contlim}

Considering $\tau$ as a small parameter, we can write the local Hamiltonian to leading order in $\tau$, 
and up to a shift for an irrelevant constant as:
\be
h(\sigma) \simeq - \tau \log \sigma^2.
\ee
The many-body map $\Psi_\tau$ can then be understood as the leading order Trotter decomposition with the time-independent Hamiltonian:
\be
H_{\rm LLL} = \lim_{\tau\to 0} \frac{1}{\tau}(H_{\rm even}+H_{\rm odd}) = 
 -\sum_{x\in\mathbb Z_N}  2\log\frac{1}{2}\left(1 + \vec{S}_x\cdot \vec{S}_{x+1}\right)
 \label{eq:HLLL}
\ee
which is, up to multiplicative and additive constants, nothing but the Hamiltonian of the isotropic LLL model \cite{Faddeev-Takhtajan}.

One can derive equations of motion in the limit $\tau\to 0$ also directly from Eqs.~(\ref{two-body-map},\ref{st}) by taking continuous-time functions
$\vec{S}_x(\tau t/2) = \vec{S}^t_x$, arriving to:
\begin{equation}
\frac{{\rm d}\mathbf{S}_x}{{\rm d}t} = 2\mathbf{S}_x \times \Big( \frac{\mathbf{S}_{x-1}}{1 + \mathbf{S}_{x-1}\cdot \mathbf{S}_x} + \frac{\mathbf{S}_{x+1}}{1 + 
\mathbf{S}_{x}\cdot \mathbf{S}_{x+1}} \Big),\quad x\in\mathbb Z_N,\; t\in\mathbb R\,.
\label{eq:LLLEOM}
\end{equation}
These are precisely the equations of motion of the LLL model, which can be equally derived as the Hamilton's equations $\dot{\vec{S}}_x = \{\vec{S}_x,H_{\rm LLL}\}$ 
for (\ref{eq:HLLL}).

Furthermore assuming that the energy is low enough so that the neighbouring spins are always almost parallel, we can approximate the spin configuration with a space-time continuous spin field
$\vec{S}(x,\tau) = \vec{S}_x(\tau)$, upon which (\ref{eq:LLLEOM}) becomes the famous integrable 
Landau-Lifshitz partial differential equation:
\be
\partial_t \vec{S} = \vec{S} \times \partial_x^2 \vec{S},\quad x,t\in\mathbb R.
\ee
The model can thus be seen as an analogue of the lattice Landau-Lifshitz model on the discrete space-time lattice. By itself this is not remarkable, since there are infinitely many symplectic space-time discretizations with such a continuous limit. What makes the above model especially interesting is its integrability, which we shall demonstrate shortly. As such, the model can be understood as the simplest integrable analogue of the LLL model in discrete time.

\subsection{Integrability of the model}
\label{sec1:integrability}
In this section we show the model is integrable, as alluded to above. The integrability of the model has been first motivated by numerically computing the 
Lyapunov spectrum of the system (not shown), which indicates that the largest Lyapunov exponent vanishes for all values of $\tau$.

The elementary two-body propagator (\ref{two-body-map}) maps a pair of 2-spheres onto itself. It clearly commutes with the permutation (transposition) mapping $\Pi$:
\begin{equation}
\Pi \circ \Phi_{\tau} = \Phi_{\tau} \circ \Pi, \quad \Pi(\mathbf{S}_1, \mathbf{S}_2) := (\mathbf{S}_2, \mathbf{S}_1) .
\end{equation}
Denoting the identity map on the 2-sphere as $\mathbb{I}$ we can extend the local dynamical map to a product of three 2-spheres, e.g.
as $(\mathbb{I} \otimes \Phi_\tau)(\vec{S},\vec{S}',\vec{S}'') = (\vec{S},\Phi_\tau(\vec{S}',\vec{S}''))$.
A lengthy but straightforward calculation proves the following remarkable identity:
\begin{equation}
\label{YBE}
\Big(\Phi_{\lambda} \otimes \mathbb{I} \Big) \circ \Big(\mathbb{I}  \otimes \Phi_{\lambda + \mu} \Big) \circ \Big( \Phi_{\mu} \otimes \mathbb{I}  \Big) = \Big( \mathbb{I} \otimes \Phi_{\mu} \Big) \circ \Big( \Phi_{\lambda+\mu} \otimes \mathbb{I}  \Big) \circ \Big( \mathbb{I} \otimes \Phi_{\lambda} \Big).
\end{equation}
Eq.~(\ref{YBE}) is the braid form of the Yang-Baxter equation for the local propagator \cite{ABA}. The model is thus a classical analogue of  the integrable trotterization of the quantum spin 1/2 chain \cite{quant_trotter}. 
Its composition with permutation $R_\lambda=\Pi\circ \Phi_\lambda$ is a baxterized solution of the
{\em set-theoretic quantum Yang Baxter equation} \cite{SetThYB1,SetThYB2} where now $\lambda\in\mathbb C$ plays the role of the spectral parameter. Note that $\Phi_0 =  \mathbb{I}^{\otimes 2}$ and $\Phi_\infty = \Pi$, whereas
$R_0 = \Pi$ and $R_\infty = \mathbb{I}^{\otimes 2}$.

While the Yang-Baxter equation is a hallmark of integrable systems and corroborates the numerical observation of zero Lyapunov exponents, we need to define
the appropriate Lax operator in order to construct the conserved quantities. We start by defining the following parameter-dependent $2\times 2$ matrix-valued function over the 2-sphere $L(\lambda) : {\cal S}^2 \to {\rm End}(\mathbb C^2)$, $\lambda\in\mathbb C$:
\be
\label{eq:Lax}
L(\vec{S};\lambda) =  \one + \frac{1}{2i \lambda} \mathbf{S} \cdot \pmb{\sigma}, \
\ee
where $\vec{\sigma} = (\sigma_1,\sigma_2,\sigma_3)$
is a vector of Pauli matrices and $\one$ is a $2\times 2$ unit matrix. By a straightforward calculation (see Appendix \ref{RLL_append} for details) we can check the following remarkable identity:
\be
L(\vec{S}_2;\lambda)L(\vec{S}_1;\mu) = L(\vec{S}'_2;\mu)L(\vec{S}'_1;\lambda),\quad (\vec{S}'_1,\vec{S}'_2) := \Phi_{\lambda-\mu}(\vec{S}_1,\vec{S}_2),
\ee
holding for any pair of {\em spectral parameters} $\lambda,\mu \in\mathbb C$,
which we recognise as a set-theoretic version of the RLL relation. We thus define a set-theoretic version of the Lax operator, by extending $L$ to a local function over the
full phase space ${\cal M}$ of classical spin-configurations:
\be
L_x(\lambda) := I^{\otimes x} \otimes L(\lambda)\otimes I^{\otimes N-x-1},\quad x\in\mathbb Z_N.
\ee
Here $I : {\cal S}^2\to\mathbb C$ is understood as a local unit function $I(\vec{S})\equiv 1$, 
while tensor products of local functions over a cartesian product of their domain sets are defined as $(f\otimes g)(\vec{x},\vec{y}) \equiv f(\vec{x})g(\vec{y})$. 
The set-theoretic RLL relation can then be written compactly without explicit reference to spin variables as:
\be
L_{x+1}(\lambda)L_{x}(\mu) = L_{x+1}(\mu)L_{x}(\lambda)  \circ \mathbb I^{\otimes x} \otimes \Phi_{\lambda-\mu}\otimes \mathbb I^{\otimes N-x-2}.
\label{RLL}
\ee
Note a small but important distinction with respect to the quantum XXX Lax operator \cite{ABA}, namely the spectral parameter $\lambda$ here (\ref{eq:Lax}) needs to be inverted in
order to become additive in the RLL relation. 

We shall now use the RLL equation to construct conserved quantities following essentially the standard procedure of the algebraic Bethe ansatz \cite{ABA}.
We define a particular, 2-parameter-dependent function over ${\cal M}$, $T(\lambda,\mu):{\cal M}\to\mathbb C$, $\lambda,\mu\in\mathbb C$, in terms of the trace
of a staggered monodromy matrix:
\bea
T(\vec{S}_0,\vec{S}_1,\ldots,\vec{S}_{N-1};\lambda,\mu) &=& 
 \mathrm{tr} \Bigg( \overleftarrow{\prod_{x=0}^{N/2-1}} L(\vec{S}_{2x+1};\lambda)L(\vec{S}_{2x};\mu) \Bigg), \nonumber \\
 {\rm or}\quad T(\lambda,\mu) &=& \mathrm{tr} \Bigg( \overleftarrow{\prod_{x=0}^{N/2-1}} L_{2x+1}(\lambda)L_{2x}(\mu) \Bigg). \label{monodromy}
\eea
Eqs.~(\ref{RLL}) and (\ref{MBmap}) immediately imply:
\bea
& T(\lambda,\mu)\circ \Psi^{\rm odd}_{\lambda-\mu} = T(\mu,\lambda),\quad
T(\mu,\lambda)\circ \Psi^{\rm even}_{\lambda-\mu} = T(\lambda,\mu),\\
& T(\lambda,\mu)\circ \eta^2 = T(\lambda,\mu),
\eea
yielding time conservation of phase-space function $T(\lambda,\lambda-\tau)$,  for any $\lambda\in\mathbb C$:
\be
T(\lambda,\lambda-\tau)\circ \Psi_\tau = T(\lambda,\lambda-\tau),
\ee
and its translational invariance for an even number of sites. $T(\lambda,\lambda-\tau)$ can thus be considered as a generating function of the system's conserved quantities, playing a role analogous to the transfer matrix in algebraic Bethe ansatz. However, conserved quantities generated by $T(\lambda,\lambda-\tau)$ will in general be complex and nonlocal in spin variables $\vec{S}_x$.
In order to proceed with defining real and local conserved quantities, we first note a few useful properties of the Lax operator which are straightforward to verify:
\begin{align}
\label{L_rels}
\overline{L_x(\lambda)} &= \sigma_2 L_x(\overline{\lambda}) \sigma_2,\nonumber\\
L_x^T(\lambda) &= \sigma_2 L_x(-\lambda) \sigma_2, \\
L(\vec{S};\mp i/2) &= |\alpha_\pm(\vec{S})\rangle \langle \beta_\pm(\vec{S})|,\qquad \langle \beta_\pm(\vec{S})|\alpha_\pm(\vec{S}')\rangle = 
1\pm \vec{S}\cdot\vec{S}', \nonumber
\end{align}
last equation meaning that at $\lambda=\mp i/2$ the Lax operator becomes a rank-1 projector, while $\overline{\bullet}$ denotes complex conjugation (in Pauli basis) assuming that $\vec{S}$ are manifestly real
variables. Using the method elaborated in \cite{Faddeev-Takhtajan} combined with the staggering of the spectral parameter,
exactly as in the  case of trotterized quantum XXX chain \cite{quant_trotter}, one then shows that the logarithmic derivatives of square-moduli of monodromies:
\bea
Q^{\rm even}_k &=& \partial_\lambda^k \log |T(\lambda,\lambda-\tau)|^2 \vert_{\lambda=-\frac{i}{2}}, \nonumber \\
Q^{\rm odd}_k &=&  \partial_\lambda^k \log |T(\lambda,\lambda-\tau)|^2 \vert_{\lambda=\tau-\frac{i}{2}},\quad k=0,1,2\ldots
\label{Qs}
\eea
form two independent sets of {\em conserved charges} which are sums of {\em local} densities:
\be
Q^{\rm even}_k = \sum_{x=0}^{N/2-1} q^{\rm even}_k \circ \eta^{2x},\quad
Q^{\rm odd}_k = \sum_{x=1}^{N/2} q^{\rm odd}_k \circ \eta^{2x-1},
\ee
where $q^{\rm even/odd}_k$ are charge densities which depend only on the first $2k+3$ spin coordinates $\vec{S}_0,\vec{S}_1\ldots\vec{S}_{2k+2}$.
We note that for $k\sim N/2$ and above the locality arguments start to break down due to periodicity in $N$, so the conserved 
quantities $Q^{\rm even/odd}_k$ are no longer independent from the previous ones (those for smaller $k$).

For concreteness, we explicitly compute the charge densities of the most local pair of conserved quantities $Q^{\rm even/odd}_0$:
\begin{align}
& {q}^{\rm odd}_{0}\circ\eta^{2x-1}
= \log  \mathrm{tr} \left(L_{2x+1}\!\left(-\frac{i}{2}\right)L_{2x}\!\left(-\tau-\frac{i}{2}\right)L_{2x-1}\!\left(-\frac{i}{2}\right)L_{2x}\!\left(\tau-\frac{i}{2}\right) \right) = \nonumber\\
&\qquad=\log \Bigg[1+ \frac{1}{1+4\tau^2} \Big(1 + 2 \mathbf{S}_{2x+1} \cdot \mathbf{S}_{2x} + 2\mathbf{S}_{2x} \cdot \mathbf{S}_{2x-1} + 4\tau^2 \mathbf{S}_{2x+1} \cdot \mathbf{S}_{2x-1} + \nonumber\\ 
&\qquad\qquad\qquad+ 2 \big(\mathbf{S}_{2x+1} \cdot \mathbf{S}_{2x} \big) \big( \mathbf{S}_{2x} \cdot \mathbf{S}_{2x-1}\big) - 4\tau \big(\mathbf{S}_{2x+1}, \mathbf{S}_{2x},\mathbf{S}_{2x-1} \big) \Big)\Bigg], \nonumber\\
& {q}^{\rm even}_{0}\circ\eta^{2x}  = 
 \log  \mathrm{tr} \left(L_{2x+1}\!\left(-\frac{i}{2}\right)L_{2x}\!\left(\tau-\frac{i}{2}\right)L_{2x-1}\!\left(-\frac{i}{2}\right)L_{2x}\!\left(-\tau-\frac{i}{2}\right) \right) =  \nonumber\\
&\qquad= \log \Bigg[1+ \frac{1}{1+4\tau^2} \Big(1 + 2 \mathbf{S}_{2x+2} \cdot \mathbf{S}_{2x+1} + 2\mathbf{S}_{2x+1} \cdot \mathbf{S}_{2x} + 4\tau^2 \mathbf{S}_{2x+2} \cdot \mathbf{S}_{2x} + \nonumber\\ 
&\qquad\qquad\qquad+ 2 \big(\mathbf{S}_{2x+2} \cdot \mathbf{S}_{2x+1} \big) \big( \mathbf{S}_{2x+1} \cdot \mathbf{S}_{2x}\big) + 4\tau \big(\mathbf{S}_{2x+2}, \mathbf{S}_{2x+1},\mathbf{S}_{2x} \big) \Big)\Bigg]. \nonumber
\end{align}
The even and odd densities differ by  a one lattice point shift and the sign of the mixed product (or sign change $\tau\to-\tau$). 

We observe that in the continuous time limit, $\tau \rightarrow 0$, the two branches of local densities converge to each other 
and the zeroth conserved quantity is precisely the Hamiltonian of LLL model:
\begin{equation}
\lim_{\tau\to 0} Q^{\rm even}_0 = \lim_{\tau\to 0} Q^{\rm odd}_0 = \sum_{x=0}^{N-1} \log \Big(2\big(1+ \mathbf{S}_x\cdot\mathbf{S}_{x+1} \big) \Big) = 
-H_{\rm LLL} + {\rm const}.
\end{equation}
To explain why Eq.~(\ref{Qs}) will generate local conserved quantities of increasingly large support we observe that the $\lambda$-derivatives of Lax matrices 
are no longer projectors at the specified values of $\lambda$. For $k\ge 1$, $Q^{\rm even/odd}_k$ are rational expressions with derivatives of Lax matrices in the numerator and the nonderived trace of monodromy matrix in the denominator. The $\langle \beta_+| \bullet |\alpha_+\rangle$ matrix element of any Lax matrix in the numerator, not adjacent to a derived Lax matrix, will be cancelled by the corresponding Lax matrix element in the denominator. This cancellation ensures the locality of the generated conserved densities. The terms with a derived Lax matrix next to an already derived Lax matrix will produce local conserved charges with an increasingly wider support. 

Since a direct computation of higher conserved charges is tedious, it would be of interest if the conserved charges of the model could be equipped with a boost operation that would facilitate their automated computation, similarly as in the quantum case \cite{quant_trotter}.

\begin{figure}[h]
\centering
\begin{tikzpicture}[baseline=(current  bounding  box.center), scale=1.1]
% Spin labels

%Left propagator
\draw[thick, fill=black] (-1, 1) circle (0.07cm); 
\draw[thick, fill=black] (-1, -1) circle (0.07cm); 
\draw[thick, fill=black] (-3, 1) circle (0.07cm); 
\draw[thick, fill=black] (-3, -1) circle (0.07cm); 

\draw[thick] (-3, -1) -- (-1, 1);
\draw[thick] (-3, 1) -- (-1, -1);

\draw[black,arrows={-Triangle[angle=90:8pt]}] (-3, -1) -- (-2.65, -0.65) ;
\draw[black,arrows={-Triangle[angle=90:8pt]}] (-1, -1) -- (-1.35, -0.65) ;

\draw[black,arrows={-Triangle[angle=90:8pt]}] (-2, 0) -- (-2.75, 0.75) ;
\draw[black,arrows={-Triangle[angle=90:8pt]}] (-2, 0) -- (-1.25, 0.75) ;

\draw[thick, fill=white, rounded corners=2pt] (-2.5, -0.5) rectangle (-1.5,0.5) node[pos=.5]{\Huge $\, \Phi_{\tau}$};
\draw[thick, fill=mygreen, fill opacity = 0.4, rounded corners=2pt] (-2.5, -0.5) rectangle (-1.5,0.5);

\Text[x=-3.3,y=-1.3]{\large $\mathbf{S}_{1}$}
\Text[x=-0.7,y=-1.3]{\large $\mathbf{S}_{2}$}
\Text[x=-3.3,y=1.3]{\large $\mathbf{S}_{1}'$}
\Text[x=-0.7,y=1.3]{\large $\mathbf{S}_{2}'$}

%Right propagator
\draw[thick, fill=black] (1, 1) circle (0.07cm); 
\draw[thick, fill=black] (1, -1) circle (0.07cm); 
\draw[thick, fill=black] (3, 1) circle (0.07cm); 
\draw[thick, fill=black] (3, -1) circle (0.07cm); 

\draw[thick] (1, -1) -- (3, 1);
\draw[thick] (1, 1) -- (3, -1);

\draw[black,arrows={-Triangle[angle=90:8pt]}] (1, 1) -- (1.35, 0.65) ;
\draw[black,arrows={-Triangle[angle=90:8pt]}] (1, -1) -- (1.35, -0.65) ;

\draw[black,arrows={-Triangle[angle=90:8pt]}] (2, 0) -- (2.75, 0.75) ;
\draw[black,arrows={-Triangle[angle=90:8pt]}] (2, 0) -- (2.75, -0.75) ;

\draw[thick, fill=white, rounded corners=2pt] (1.5, -0.5) rectangle (2.5,0.5) node[pos=.5]{\Huge $\, \tilde{\Phi}_{\tau}$};
\draw[thick, fill=myred, fill opacity = 0.4, rounded corners=2pt] (1.5, -0.5) rectangle (2.5,0.5);

\Text[x=3.3,y=-1.3]{\large $\mathbf{S}_{2}$}
\Text[x=0.7,y=-1.3]{\large $\mathbf{S}_{1}$}
\Text[x=3.3,y=1.3]{\large $\mathbf{S}_{2}'$}
\Text[x=0.7,y=1.3]{\large $\mathbf{S}_{1}'$}

\end{tikzpicture}
  \caption{Definition of the dual 2-spin symplectic mapping $\tilde{\Phi}_\tau$ by a simple relabelling of the domain (in-to-box arrows) and image (out-of-box arrows) argument pairs.}
\label{fig:FigSTDual}
\end{figure}
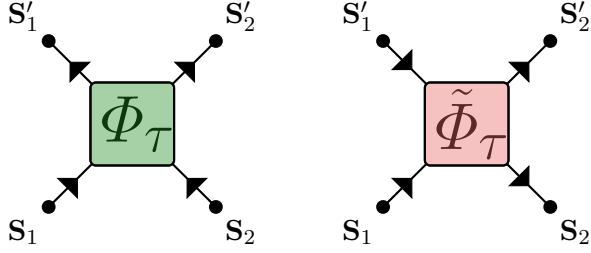

\subsection{Space-time self-duality}
The two-body propagator (\ref{two-body-map}) specifies the temporal dynamics. Knowing the values of two adjacent spins at the same time allows us to compute their time evolution at a latter time:
\begin{equation}
(\mathbf{S}_1, \mathbf{S}_2) \xrightarrow{\Phi_{\tau}} (\mathbf{S}'_1, \mathbf{S}'_2).
\end{equation}
We are interested in the existence of a dual spatial propagator, that would allow us to uniquely evolve any pair of temporally consequent spins with the same spatial coordinate to a consequent pair at one spatial point to the right, i.e. we are interested in the existence of a mapping
$\tilde{\Phi}_\tau : {\cal S}^2\times {\cal S}^2\to {\cal S}^2\times {\cal S}^2$ (depicted schematically in Fig.~\ref{fig:FigSTDual}):
\begin{equation}
(\mathbf{S}_1, \mathbf{S}'_1) \xrightarrow{\tilde{\Phi}_\tau} (\mathbf{S}_2, \mathbf{S}'_2).
\end{equation}   
Generic 2-spin symplectic maps would of course not have such a dual space-propagator, the map would typically be either not unique (non-deterministic)
or would not be defined for some pairs of spins $(\vec{S}_1,\vec{S}_1')$.
Remarkably, the two-body propagator (\ref{two-body-map}) admits such a 
dual propagator which is a bijection on ${\cal S}^2\times {\cal S}^2$, as is shown in Appendix \ref{self-dual_append}:
\bea
\label{two-body-space-map}
\tilde{\Phi}_{\tau}  (\mathbf{S}_1, \mathbf{S}'_1) &=&  \frac{1}{\delta^2 + \tau^2} \Big(\!-\!\delta^2\mathbf{S}_1 + \tau^2\mathbf{S}'_1 - 
\tau \mathbf{S}_1\times\mathbf{S}'_1,-\delta^2\mathbf{S}'_1 + \tau^2\mathbf{S}_1 - \tau \mathbf{S}_1\times\mathbf{S}'_1 \Big),\quad \nonumber\\
\delta^2 &:=& \frac{1}{2} \Big(1 - \mathbf{S}_1\cdot \mathbf{S}'_1 \Big).
\eea
Moreover, the spatial dynamics have, after a simple local gauge transformation of the lattice, exactly the same form as the temporal dynamics. 
Specifically, on the level of two-body propagators, we find the following gauge equivalences between
(\ref{two-body-map}) and (\ref{two-body-space-map}):
\be
\Xi\circ \tilde{\Phi}_\tau = \Phi_\tau\circ(-\Xi), \qquad (-\Xi)\circ\tilde{\Phi}_{\tau} = \Phi_{-\tau}\circ \Xi, \label{GE}
\ee
where $\Xi(\vec{S},\vec{S}') := (\vec{S},-\vec{S}')$,  $(-\Xi)(\vec{S},\vec{S}') := (-\vec{S},\vec{S}')$.
This means that flipping the signs of the spins on the checker-board pattern:
\begin{equation}
\label{dual_trans0}
\tilde{\mathbf{S}}^t_x = (-1)^{x+t+1} \mathbf{S}^t_x,
\end{equation}
the {\em spatial} dynamics are given by the {\em temporal} two-body propagator (\ref{two-body-map}) (see Fig.~\ref{fig:trotter}):
\be
(\tilde{\vec{S}}^{2t}_{2x+1},\tilde{\vec{S}}^{2t+1}_{2x+1}) = \Phi_\tau(\tilde{\vec{S}}^{2t}_{2x},\tilde{\vec{S}}^{2t+1}_{2x}), \qquad
(\tilde{\vec{S}}^{2t-1}_{2x+2},\tilde{\vec{S}}^{2t}_{2x+2}) = \Phi_\tau(\tilde{\vec{S}}^{2t-1}_{2x+1},\tilde{\vec{S}}^{2t}_{2x+1}).
\label{stdual}
\ee
Since the forms of the temporal dynamics and its dual spatial dynamics coincide, up to a local sign gauge, the model is said to be {\em space-time self-dual}.
Likewise, flipping the signs of the spins along the complementary checker-board:
\begin{equation}
\mathbf{\tilde{S}}^t_x = (-1)^{x+t} \mathbf{S}^t_x,
\label{compl}
\end{equation}
gives the the same spatial dynamics (\ref{stdual}), but with the opposite value of $\tau$.
This concludes the preliminary analytical investigation of the model.

We note that space-time duality has been discussed in space-time continuous integrable field theories, where existence of a unique space dynamics can be connected to a Lax zero-curvature condition \cite{Doikou}. However, in discrete space-time setting it is not clear if space-time (self-)duality is connected to integrability, in particular, since in quantum systems it has been found even in maximally chaotic dynamics \cite{Bruno,Bruno1,Sarang}.

\section{Correlation functions}
In order to determine the dynamical properties of the model we numerically compute the connected spin-spin spatio-temporal autocorrelation function defined as:
\begin{equation}
\hat{C}(x, t) = \langle S^{t}_{x} S^0_{0} \rangle - \langle S^{t}_{x} \rangle \langle S^0_{0} \rangle,\quad r\in\mathbb Z_{N},\; t\in\mathbb Z,
\label{eq:cf}
\end{equation}
where $S^t_x = \vec{e}\cdot \vec{S}^t_x$ is a fixed (say $z$) component of the spin ($\vec{e}=(0,0,1)$) and $\langle . \rangle$ denotes the 
average in an equilibrium ensemble (state). The equilibrium state should be invariant under time, space, and diagonal translations, 
$(x,t)\to (x,t+2)$, $(x,t)\to (x+2,t)$, and $(x,t)\to (x+1,t+1)$, respectively (c.f. symmetries of the space-time lattice depicted in Fig.~\ref{fig:trotter}).
This means, that for any observable -- function $A \in L^1({\mathcal M})$, we have the following identities:
\be
\ave{A} = \ave{A\circ \eta^2} = \ave{A \circ \Psi_\tau} = \ave{A\circ \eta \circ \Psi^{\rm even}_\tau} = \ave{A \circ \Psi^{\rm odd}_\tau\circ \eta}.
\ee
As a consequence, one point can always be shifted to $(0,0)$ in the 2-point correlation function (\ref{eq:cf}) which only depends on the difference of space and time coordinates.

\subsection{Maximum entropy probability distribution}
The equilibrium states used in computations $\ave{A}=\int\rho A\prod_x {\rm d}\vec{S}_x$, given in terms of a probability distribution $\rho$ over ${\cal M}$,
are assumed to be separable and translationally invariant, implying that $\rho= \prod_{x} \rho(\vec{S}_x)$, i.e. each spin is independently, identically distributed.
Fixing an average value of a component $S_3=\vec{e}\cdot\vec{S}$ of magnetization, $\mu=\ave{S_{3}}$, we seek for $\rho(\vec{S})\equiv \rho_\mu(\varphi,\vartheta)$ written in  spherical coordinates in terms of a polar and azimuthal angles $(\varphi,\vartheta)$,
which maximizes the entropy $\Sigma$:
\begin{align}
\Sigma &= - \int_{-1}^{1}\!{\rm d}(\cos \vartheta)\int_0^{2\pi}\!{\rm  d}\varphi\,\rho_\mu(\varphi,\vartheta) \log \rho_\mu(\varphi,\vartheta) ,\\
\mu &=   \int_{-1}^{1}\!{\rm d}(\cos \vartheta)\int_0^{2\pi}\!{\rm  d}\varphi\,\rho_\mu(\varphi,\vartheta) \cos \vartheta.
\end{align}
Solving the corresponding Euler-Lagrange equations and taking into account the normalization of the probability distribution we find an explicit probability density for each spin:
\begin{equation}
\label{distribution}
\rho_{\mu} (\varphi,\vartheta) = \frac{1}{4\pi} \frac{\kappa(\mu)}{ \sinh \kappa(\mu)} e^{\kappa(\mu) \cos \vartheta}, \quad \coth \kappa(\mu) - \frac{1}{\kappa(\mu)} = \mu.
\end{equation}
This state can be interpreted as a magnetic grand-canonical equilibrium for a magnetization conserving classical spin chain.

Therefore, we numerically compute $\hat{C}(x,t)$ (\ref{eq:cf}) by Monte Carlo sampling over $M$ initial spin configurations $(\vec{S}^0_0,\vec{S}^0_1,\ldots,\vec{S}^0_{N-1})$, where each spin $\vec{S}^0_x$ is sampled
from probability distribution (\ref{distribution}), and then estimating the correlator as:
\be
\hat{C}(x,t) = \frac{1}{M}\!\!\sum_{(\vec{S}^0_0,\vec{S}^0_1,\ldots,\vec{S}^0_{N-1})}\!\!\frac{2}{(t_{\rm max}-t+1)N} \sum_{t'=0}^{t_{\rm max}-t}\sum_{x'=0}^{N/2-1} S^{t+2t'}_{x+2x'} S^{2t'}_{2x'} - \mu^2,
\ee
where we noted that $\ave{S^t_x}=\mu$. 
The above sum can be efficiently computed using the convolution theorem. The maximal number of time steps in all simulations is equal to half the number of lattice sites, $t_{\rm max}=N/2$, 
to exclude any possible artefacts due to periodic boundary conditions. 

In order to avoid even-odd (staggering) effects we in the following analyze the auto-correlation function of local magnetization averaged over a pair of neighbouring spins $A_x = \frac{1}{2}(S_x + S_{x+1})$ propagated for integer multiples of
Floquet period:
\be
C(x,t) = \ave{(A_x \circ \Psi^t_\tau) A_0} - \mu^2 =  \frac{1}{2}\hat{C}(x,2t) + \frac{1}{4}\hat{C}(x-1,2t) + \frac{1}{4}\hat{C}(x+1,2t).
\label{AA}
\ee

\subsection{Non-magnetized state ($\mu=0$)}

\begin{figure}
 
  \centering
  \includegraphics[width=\linewidth]{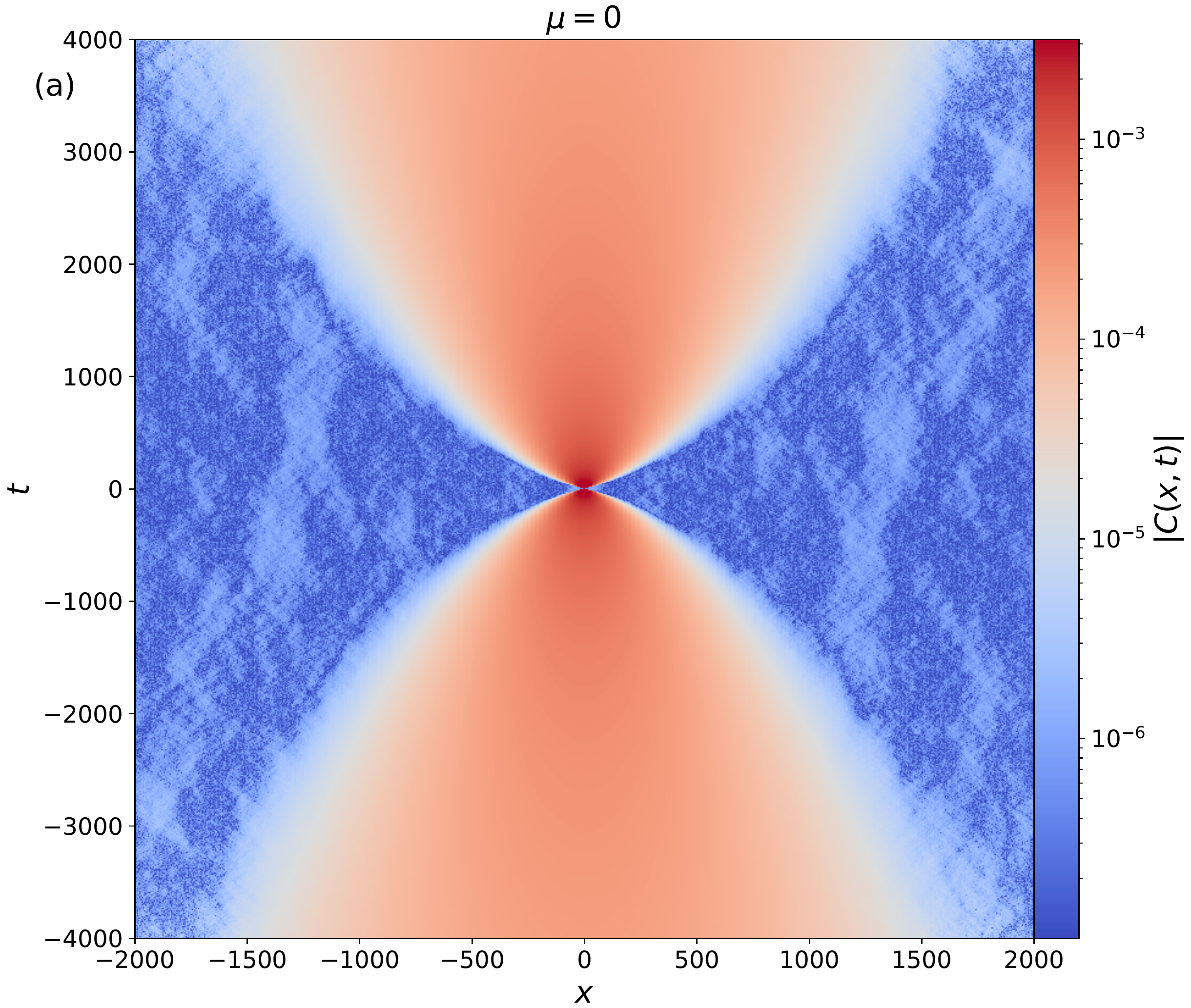}
  
  \smallskip
  
  \includegraphics[width=\linewidth]{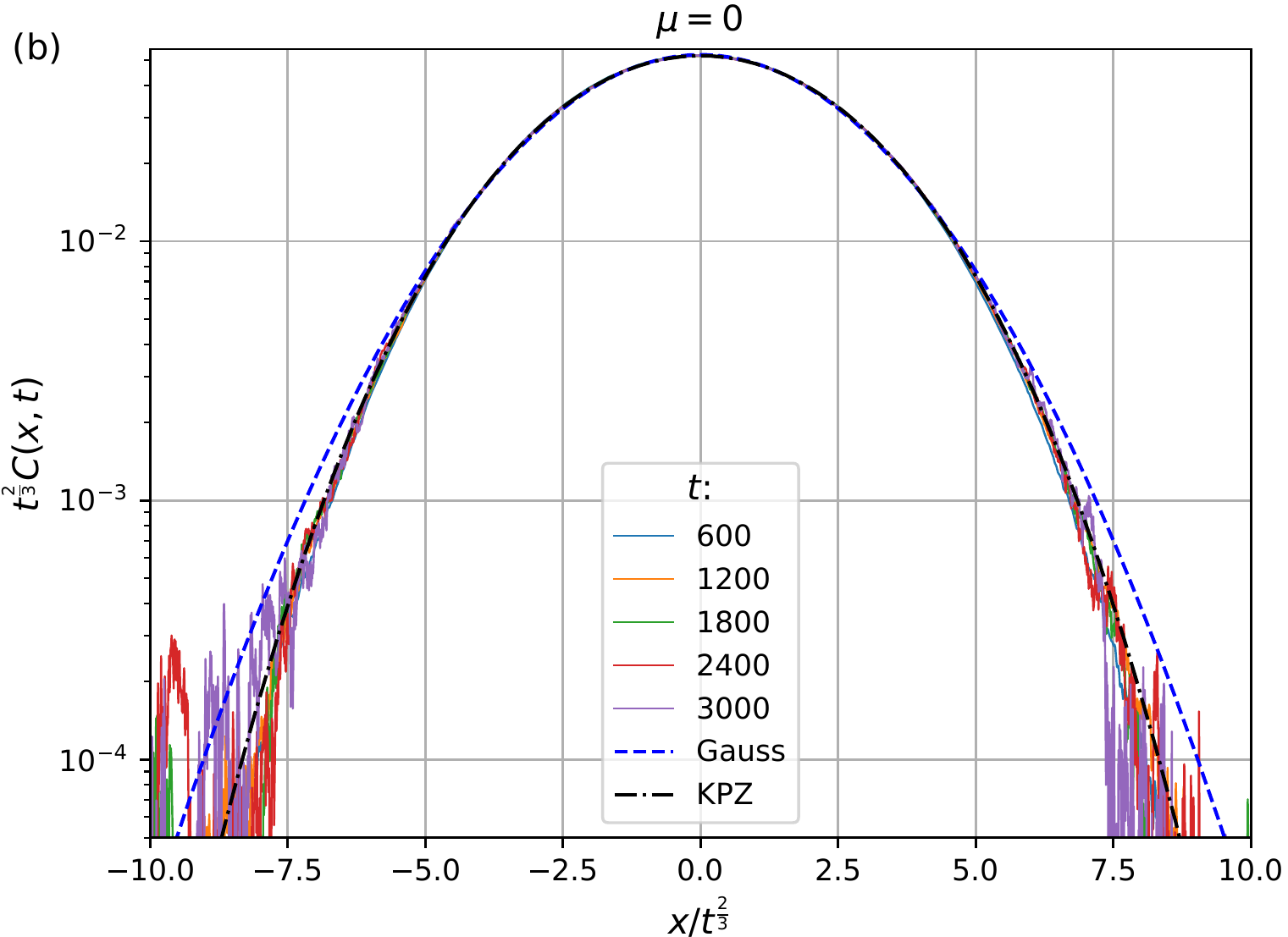}

\caption{Spin autocorrelation function (\ref{AA}) of the integrable space-time discrete dynamics (\ref{st}) at zero average magnetization $\mu=0$. Parameters of the simulation: $\tau = 1$, number of lattice sites $N= 2^{13}$, final simulation time $t_{\rm max}= 2^{12}$, averaging over a sample of $M=10^5$ initial spin configurations sampled from distribution (\ref{distribution}). Panel (a) shows a density plot of $|C(x,t)|$ in log-scale indicated in the legend. In (b) we plot snapshots of autocorrelation function cross sections rescaled according to (\ref{KPZscale}), at times indicated in the legend. The dotted lines show the best-fit Gaussian and KPZ scaling functions $g_{\rm PS}$ \cite{KPZscaling}. The KPZ fit is of the form $C(x,t)t^{\frac{2}{3}} = a g \Big(b x/t^{\frac{2}{3}}\Big)$, with $a =  0.024$, $b = 0.29$. 
}
\label{fig:test}
\end{figure}

The autocorrelation function for zero average magnetization, $\mu=0$, is shown in Figure~\ref{fig:test}a. The central (`heat') peak, spreading sub-ballistically is clearly pronounced, while no moving ('sound') peaks
can be detected. To characterize the rate of the spreading of the heat peak we rescale the late-time cross sections of the correlation function as:
\begin{equation}
\label{KPZscale}
\tilde{C}( x/t^{2/3}, t) = t^{2/3} C(x,t),
\end{equation}
whereupon we obtain a stationary profile, i.e. $\tilde{C}(\xi,t)$ is independent of $t$ for $t\gg 1$, as clearly demonstrated in Figure~\ref{fig:test}b. 
The large size of the simulated system allows the computation of the correlation function with the accuracy around three orders of magnitude requisite to 
distinguish between Gaussian and KPZ scaling \cite{KPZ}. The scaling is found to be well described by the KPZ scaling function $g_{\rm PS}(\xi)$ computed by Pr\" ahofer and Spohn \cite{KPZscaling}:
\be
\lim_{t\to\infty} \tilde{C}(\xi,t) = a g_{\rm PS}(b \xi),\quad a\approx 0.024,\; b\approx 0.29,
\ee
in excellent agreement with simulations in a related continuous time LLL model \cite{DharSpohn2} (while a similar scaling exponent was already observed in \cite{LLL_numerics}).

\subsection{Magnetized states ($\mu\neq 0$)}

\begin{figure}[h]
\centering
  \includegraphics[width=.49\linewidth]{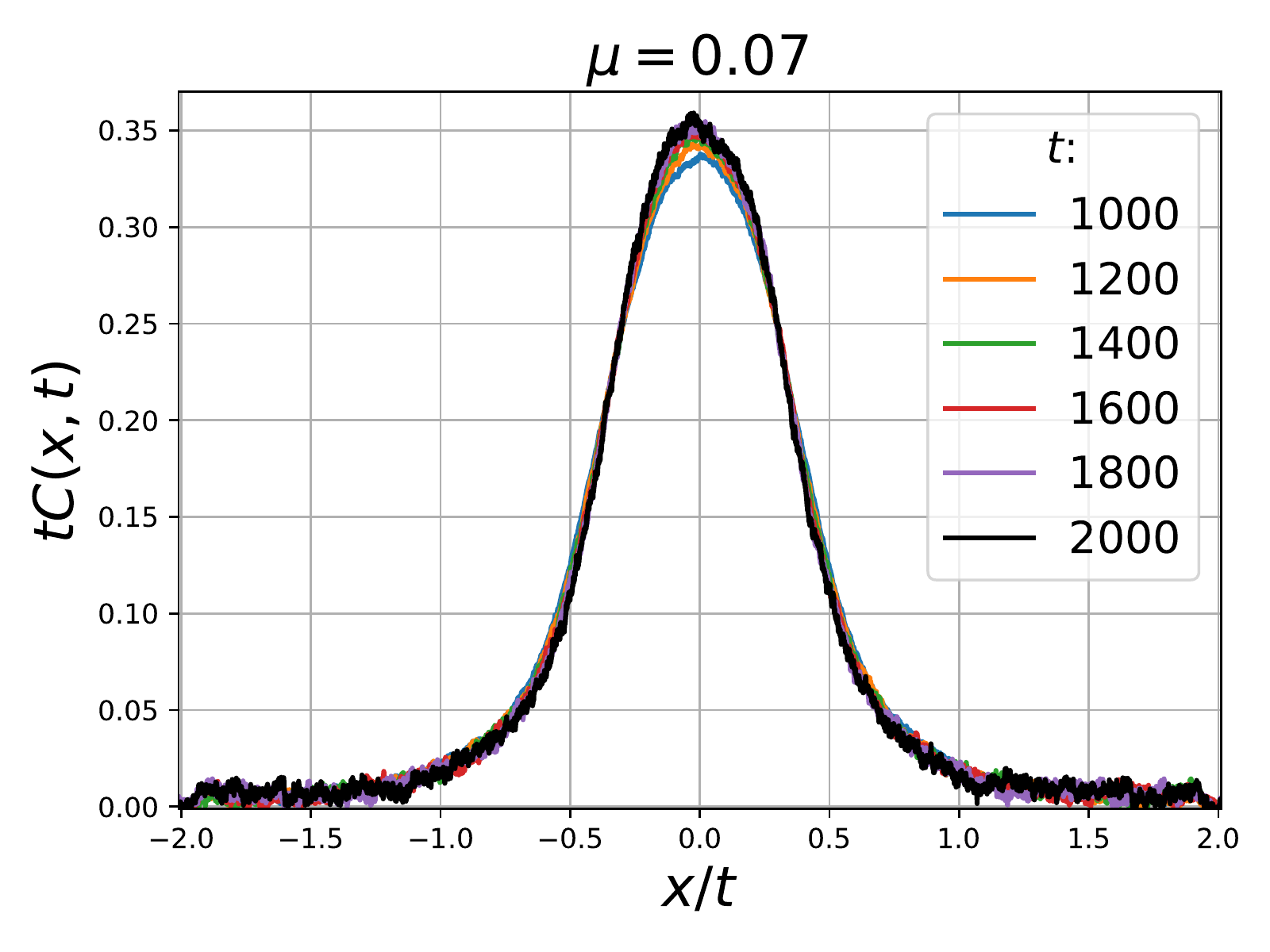}
  \includegraphics[width=.49\linewidth]{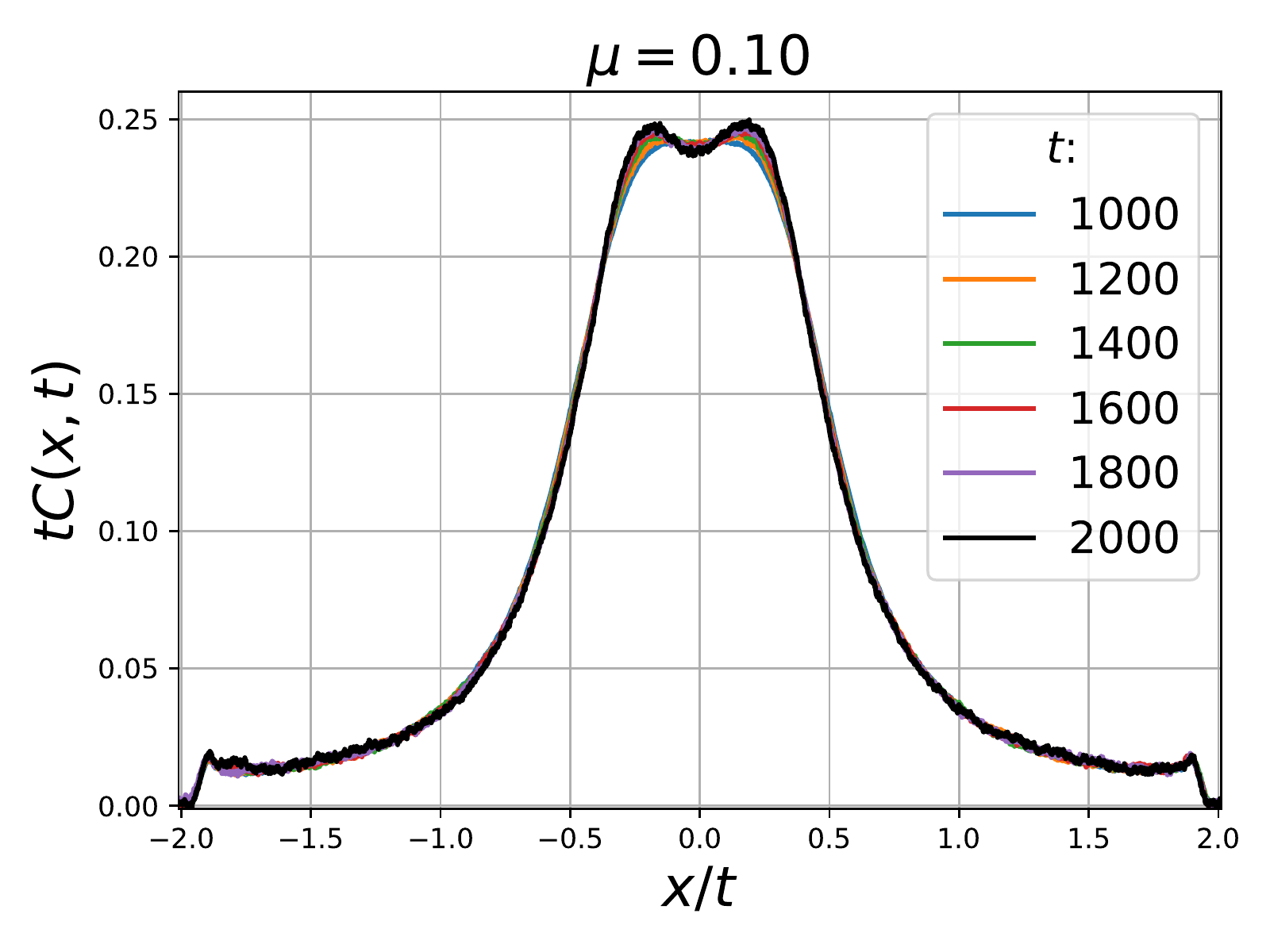}
  \includegraphics[width=.49\linewidth]{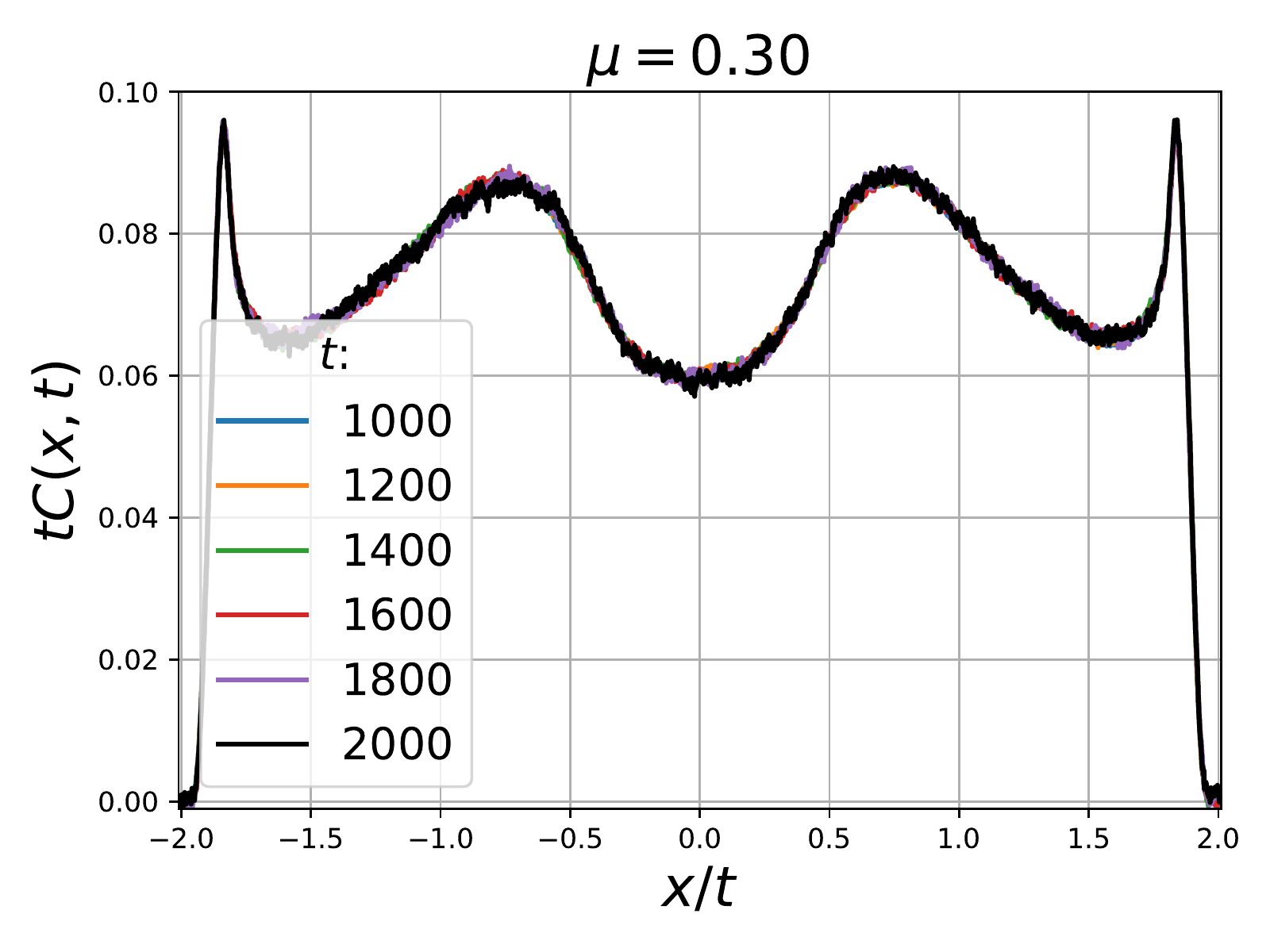}
  \includegraphics[width=.49\linewidth]{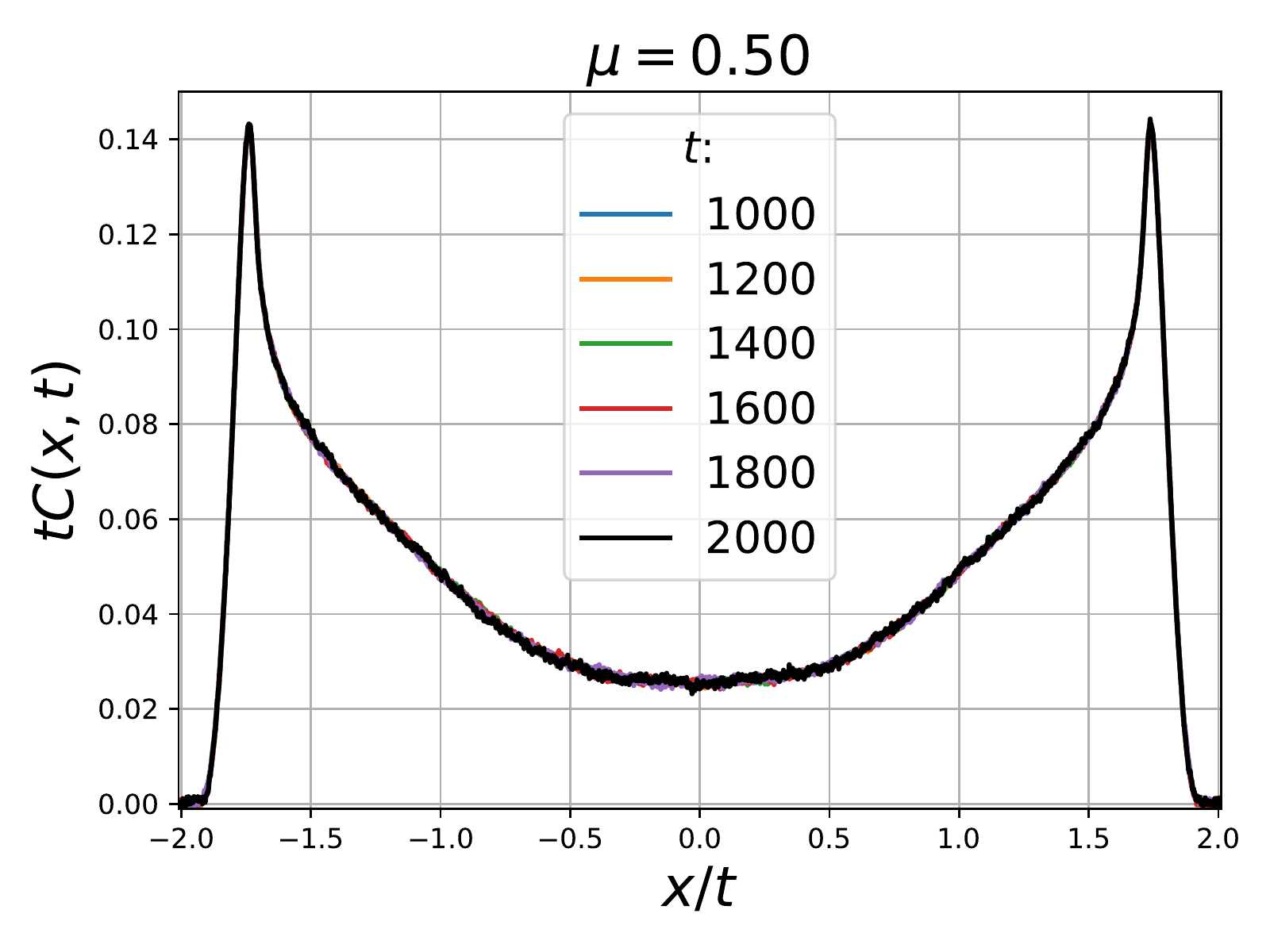}
  \includegraphics[width=.49\linewidth]{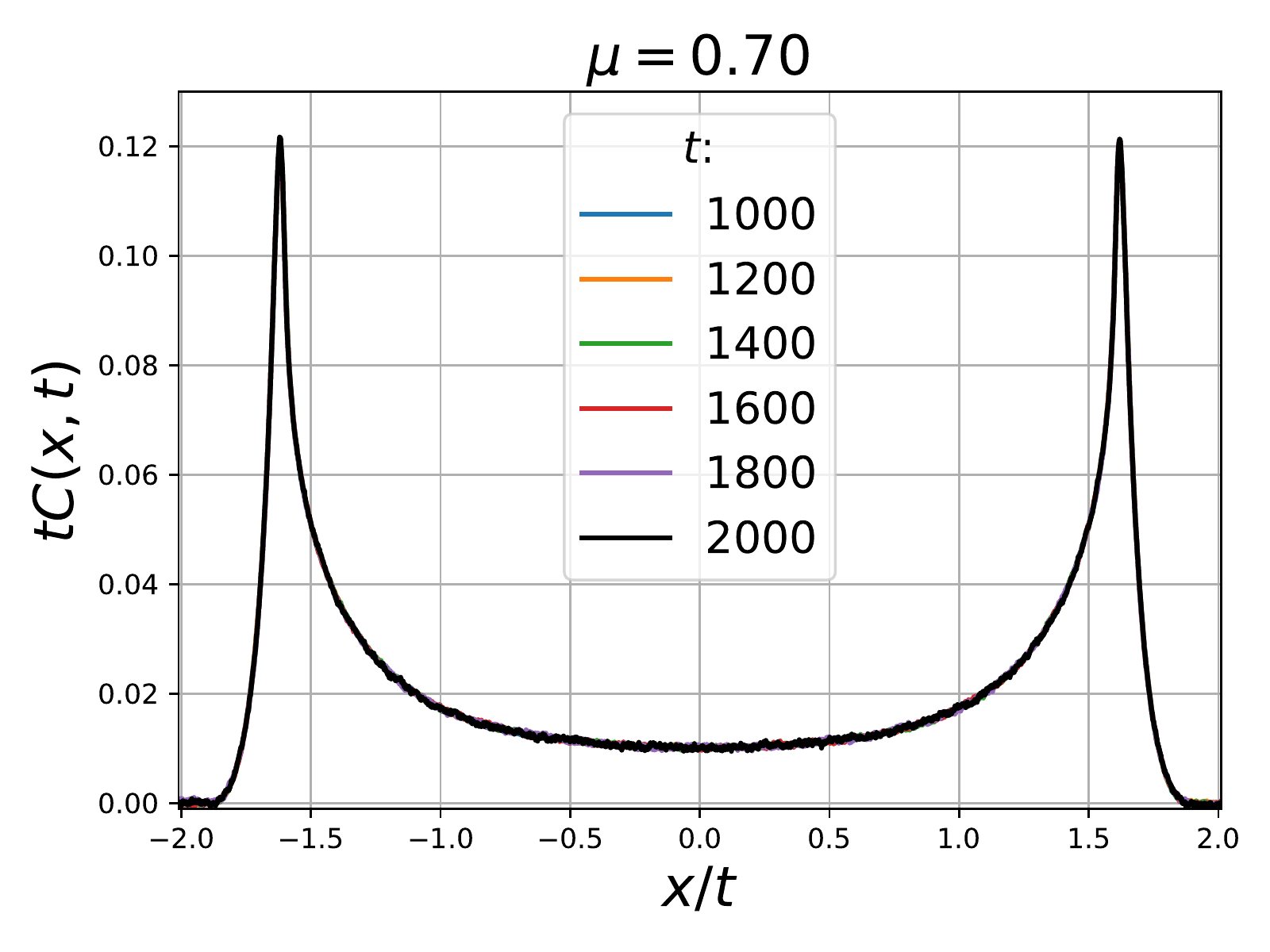}
  \includegraphics[width=.49\linewidth]{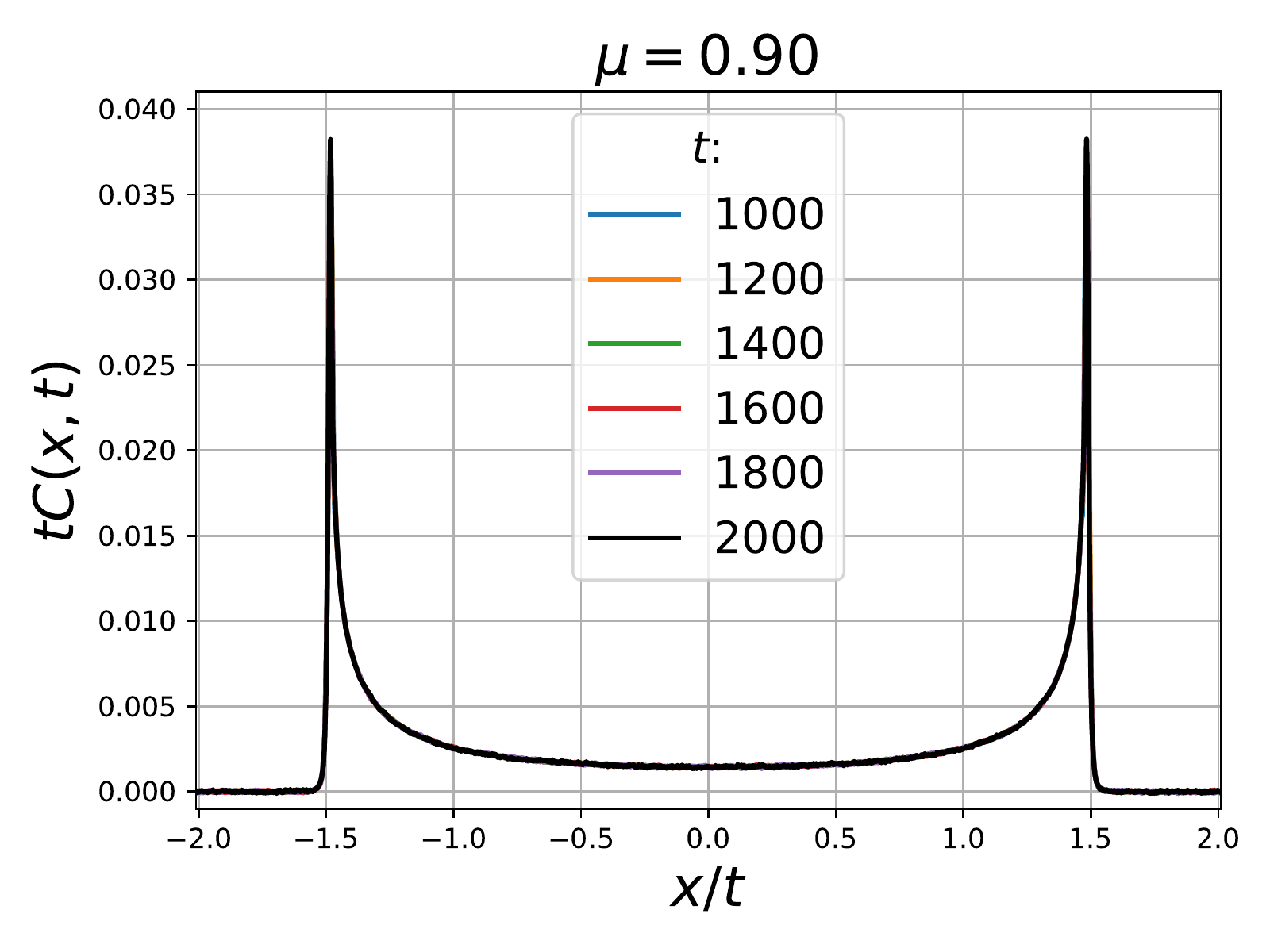}
\caption{Rescaled spin-spin correlation functions of magnitized states, stationary under a ballistic rescaling (\ref{balscale}). Edge contributions increase with growing magnetization. Initial conditions drawn from distribution (\ref{distribution}). Parameters of the simulation: $\tau = 1$, number of lattice sites $N = 2^{13}$, simulation time $t_{\rm max}= 2^{12}$, Monte Carlo samples $M = 10^5$, for average magnetizations $\mu \in \{0.07, 0.1, 0.3,  0.5, 0.7, 0.9\}$ indicated on top of each panel.
}
\label{fig:Fig3}
\end{figure}

At nonzero average magnetizations $\mu$ the spin correlation functions are no longer described by KPZ scaling (\ref{KPZscale}). Instead, a ballistic scaling:
\begin{equation}
\label{balscale}
\tilde{C}(x/t, t) = t C(x, t),
\end{equation}
gives stationary cross section, i.e. limits $\lim_{t\to\infty}\tilde{C}(\xi,t)$ quickly converge as seen in Figure~\ref{fig:Fig3}.

At very small magnetizations $\mu$ the central peak remains mostly intact, and then slowly starts splitting into two peaks at moderate $\mu$.  Increasing $\mu$ leads to growing edge contributions at around $1.4< x/t <2 $, which dominate the dynamics at high magnetizations $\mu\sim 1$. 
Note that the dominant part of the spin-spin correlations can be qualitatively interpreted as spin-wave excitations even considerably away from the fully polarized state $\mu=1$. 
The rate of convergence to the ballistic stationary state decreases with decreasing $\mu$, as can be seen in the upper-left  of Figure~\ref{fig:Fig3}, due to a crossover to KPZ scaling. This simply means that the limits $\mu\to 0$ and $t\to\infty$, both taken after the thermodynamic limit $N\to\infty$, cannot be exchanged. It would be interesting to see if the double-scaling ansatz $\mu\to 0$, $t\to\infty$, which has been proposed for quantum XXX model \cite{Romain2} would be applicable here as well, but this would require much more refined simulations on this particular regime which are beyond the scope of the present work.
A quantitative theoretical explanation of stationary correlation cross sections displayed in Figure~\ref{fig:Fig3} is within the scope of generalized hydrodynamics of classical integrable systems \cite{gen_hydro}, since $C(x,t)$ can be related to an inhomogeneous quench problem for a step initial state in the linear response limit \cite{quant-KPZ}, provided one could facilitate local conserved charges constructed in section \ref{sec1:integrability}. This is an interesting problem for future research.

\begin{figure}[h]
\centering
\begin{subfigure}{.5\textwidth}
  \centering
  \includegraphics[width=\linewidth]{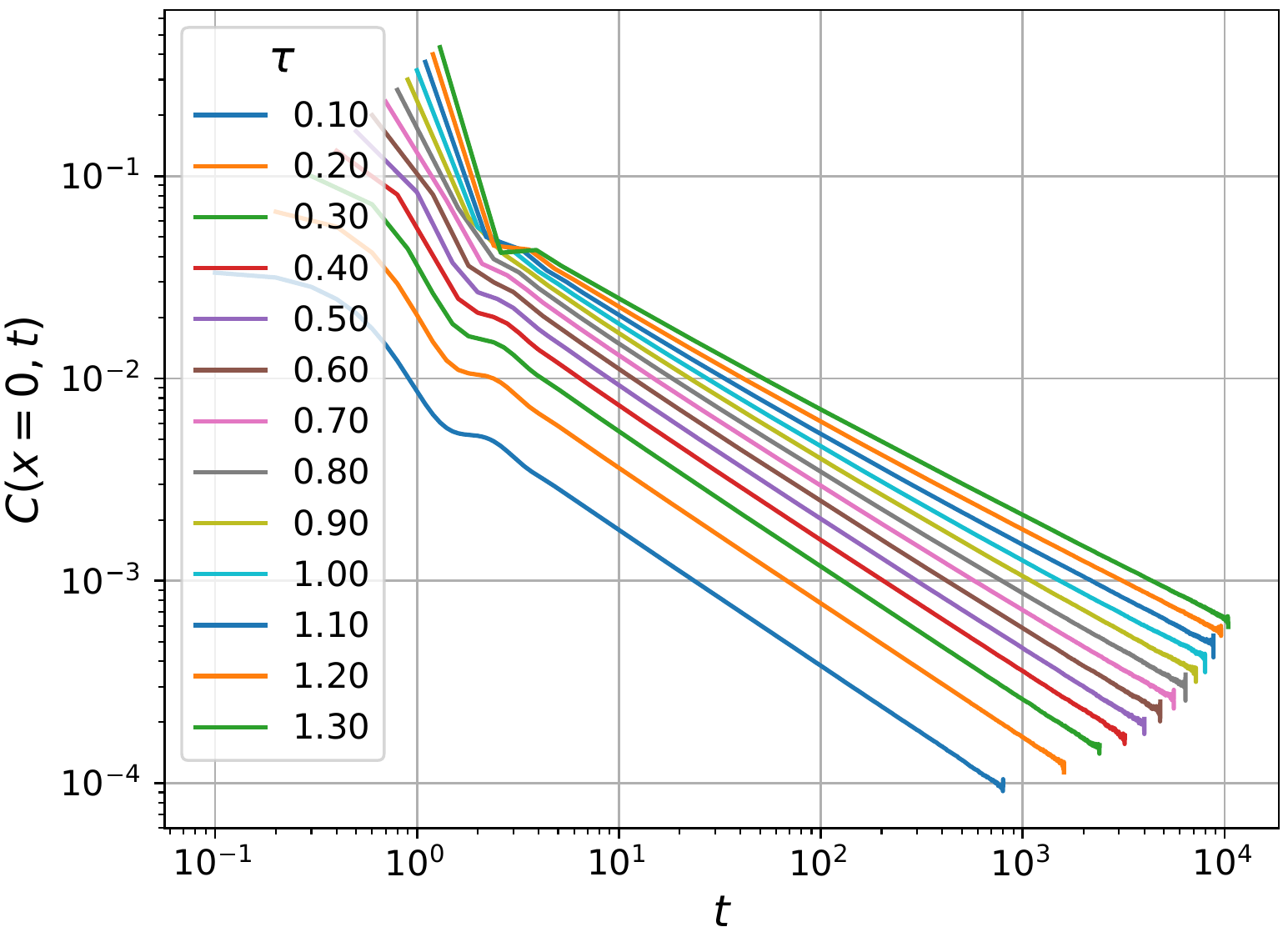}
  \caption{}
  \label{fig:Fig4left}
\end{subfigure}%
\begin{subfigure}{.5\textwidth}
  \centering
  \includegraphics[width=\linewidth]{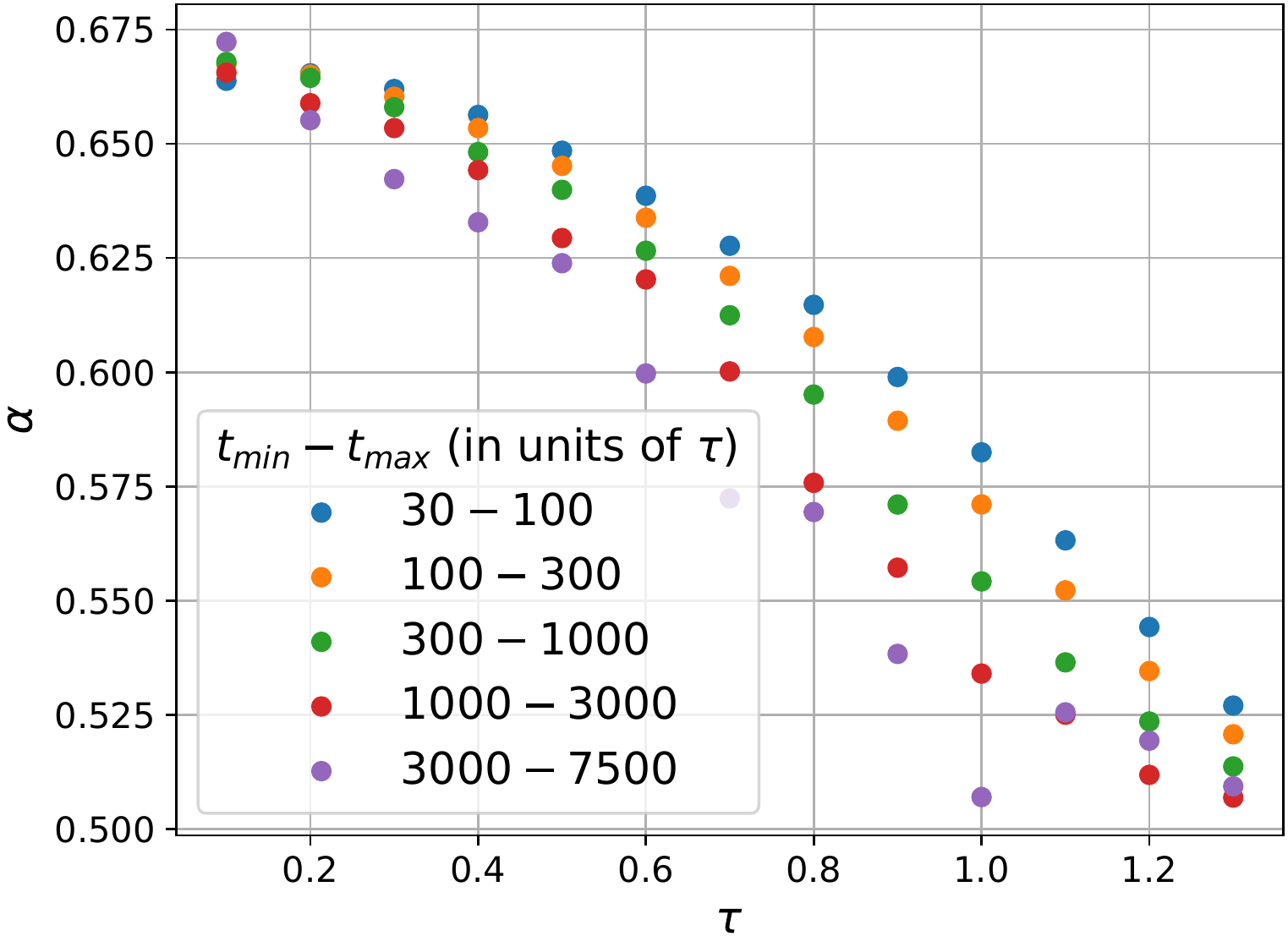}
  \caption{}
  \label{fig:Fig4right}
\end{subfigure}
\caption{Local spin autocorrelation function of the non-integrable trotterization of LLL model (\ref{nintTrotter}) in a non-magnetized state $\mu=0$. Parameters of the simulation: lattice size $N= 2^{13}$, maximal simulation time $t_{\rm max}= 2^{12}$, and averaging over $M=10^5$ initial states sampled from the distribution (\ref{distribution}). 
In (a) we show $C(x=0,t)$ vs time $t$ for various values of parameter $\tau$  which clearly suggests power-law scaling $\sim t^{-\alpha}$ after an initial transient.
A close inspection reveals that the log-log scale lines are slightly curved indicating a slow drift of the dynamical exponent.
In (b) we plot a local power law exponent $\alpha$ fitted within different time windows of geometrically scaling widths as indicated in the legend.
A systematic convergence with time towards $\alpha \approx \frac{1}{2}$ is observed.}
\label{fig:Fig4}
\end{figure}

\subsection{Nonintegrable trotterization of lattice Landau-Lifshifz model}
As we have demonstrated in the previous section an integrable trotterization of the LLL model belongs to the KPZ universality class even for a large value of time-step parameter $\tau\sim 1$, 
which which is consistent with observations in integrable quantum XXX spin $1/2$ chain and its integrable trotterization \cite{quant-KPZ}.
We note however, that the integrable trotterization of the XXX model is generated by the same local terms as the total XXX Hamiltonian.
 
Our classical integrable discrete time model, on the other hand, is generated by a different local hamiltonian (\ref{eq:localh}), which only reduces to a local LLL hamiltonian in the small $\tau$ limit.
The standard symplectic trotterization of LLL dynamics in which generators of two-step protocols (\ref{eq:two-step}) are divided into even and odd pairs: \begin{align}
\label{nintTrotter}
H^{\rm even}_{\rm LLL} &= \sum_{x=0}^{N/2-1} h_{2xj, 2x+1}, \quad H^{\rm odd}_{\rm LLL} = \sum_{x=1}^{N/2} h_{2x-1, 2x}\\
& h_{x, x+1} = \log \Big(1+ \mathbf{S}_{x} \cdot \mathbf{S}_{x+1}\Big), \nonumber
\end{align}
is {\em non-integrable}, with integrability breaking terms being of order ${\cal O}(\tau^2)$.  Computation of the full Lyapunov spectrum for various $\tau$ indeed confirmed chaoticity and hence non-integrability of the model. 
Checking the scaling of dynamical correlation functions of this trotterized LLL model would therefore be a stringent test of condition of complete integrability for the emergence of KPZ scaling.

The most accurate numerical information is the height of the peak of the correlation function (or equal-space correlator) which decays with the dynamical exponent  $C(x=0,t) \propto t^{-\alpha}$, with $\alpha=1/z$.
We have observed that precise determination of $\alpha$ for small values of $\tau$ is numerically very difficult. Fitting power laws to $C(0,t)$, i.e. linear fit of $\alpha,\beta$ to $\log C(0,t)=-\alpha \log t+\beta$, for different windows of  time $t$ 
we have indeed confirmed that, for any $\tau \neq 0$, the running exponent $\alpha$ moves towards $1/2$ (signalling $z\to 2$) by increasing $t$ (see Figure~\ref{fig:Fig4}). 
A clear convergence could only be achieved for $\tau > 0.8$, however, our results are not inconsistent with the conclusion that the SO(3) symmetric trotterized LLL model is diffusive, except in the integrable point $\tau = 0$ where it exhibits KPZ scaling.

We note that the traces of weakly broken integrability can impact the model's dynamics across several orders of magnitude in time, with normal diffusion predominating only on very long timescales. 
The above results indicate that the integrability of the model is indeed a necessary condition for the appearance of the KPZ scaling.

\section{Discussion and conclusions}

We have constructed a new integrable two-dimensional discrete and deterministic space-time lattice dynamics in terms of a set theoretic baxterized R-matrix.\footnote{It is remarkable that integrability of this classical discrete-time model is related to the quantum Yang-Baxter equation, while its continuous-time limit (the LLL model) is related to the classical r-matrix satisfying the classical Yang-Baxter equation \cite{Faddeev-Takhtajan}. ``Quantum-classical'' correspondence in this context seems to be equivalent to a discrete-continous time transition.} The latter can be expressed in terms of a 4-dimensional rational symplectic map with manifest rotational (SO(3)) symmetry. 
The map can be generated by a simple 2-particle (or 2-spin) hamiltonian and reduces to LLL model in the continuous time limit, thus it can be interpreted in two ways: either as an integrable symplectic many-body map with local interactions (or integrable classical Floquet circuit), or as an integrable time-discretization (or trotterization) of an integrable LLL model. Writing a $2\times 2$ set theoretic Lax matrix, satisfying the corresponding RLL relation, we construct an extensive
family of conserved local charges. Numerical simulations of dynamical spin-spin correlation functions reveal a clear KPZ scaling in the unconstrained maximum-entropy state with vanishing average magnetization, or the ballistic scaling in a magnetized state.
While the former calls for further theoretical understanding, the latter is to be expected based on theory of generalized hydrodynamics in integrable systems.

One hint for a theoretical analysis of the KPZ scaling in such and related models might come from a curious space-time duality symmetry of the model, 
namely dynamics can be propagated deterministically and reversibly in space direction as well, using the same 2-particle symplectic map. This may lead to in intimate connection between equal-time and equal-space correlators which might yield KPZ scaling in equal-space correlators as one of the self-consistent solutions. This is an interesting topic for further research.

Another question which naturally appears is the possibility of $q-$deformation of our symplectic mapping and construction of the corresponding easy-plane and easy-axis spin-lattice dynamics which -- according to the standard wisdom (see e.g.~\cite{LLL_numerics}) -- should have
ballistic and diffusive dynamics at vanishing magnetization, and a nontrivial spin Drude weight in the former regime.

Curiously, the numerical accuracy of dynamical correlation functions that can be obtained within a moderate computation time is hardly comparable to the accuracy of the corresponding simulation of the quantum XXX chain, which should -- according to naive expectations -- be much harder. The reason is probably in the fact that for a specific initial mixed state (namely high-temperature mixed state with weak magnetization bias) the simulation of dynamics of quantum density matrix is very efficient. It would be desirable to investigate if a similar classical simulation could be performed in the Liouville picture, representing the statistical ensemble (joint probability distribution of classical spin chain) in terms of a {\em matrix product ansatz}.

\begin{acknowledgements}
We acknowledge fruitful discussions with J. De Nardis, E. Ilievski, K. Klobas, M. Medenjak, V. Popkov, H. Spohn, M.~Vanicat, M.~Ljubotina and L.~Zadnik. This work has been supported by the European Research Council under the Advanced Grant No. 694544 -- OMNES, and by the Slovenian Research Agency (ARRS) under the Programme P1-0402. 
\end{acknowledgements}




\appendix

\section{appendix: Proof of RLL relation}
\label{RLL_append}
Central to the proof of integrability of the model is the RLL relation:
\be
L(\vec{S}_2;\lambda)L(\vec{S}_1;\mu) = L(\vec{S}'_2;\mu)L(\vec{S}'_1;\lambda),\quad (\vec{S}'_1,\vec{S}'_2) := \Phi_{\lambda-\mu}(\vec{S}_1,\vec{S}_2),
\label{RLL_app}
\ee
which we shall verify explicitly below.

The Lax matrix has been defined as:
\begin{equation}
L(\vec{S}; \lambda) = \one + \frac{1}{2i \lambda} \vec{S}_n \cdot \pmb{\sigma},
\end{equation}
where $\vec{S}$ lies on the 2-sphere and $\pmb{\sigma}$ is a vector of Pauli matrices:
\be
\pmb{\sigma} = (\sigma_1, \sigma_2, \sigma_3).
\ee
The two-body propagator is defined as:
\begin{align}
&\Phi_{\tau}  (\mathbf{S}_1, \mathbf{S}_2) =  \frac{1}{\sigma^2 + \tau^2} \Big(\sigma^2\mathbf{S}_1 + \tau^2\mathbf{S}_2 + \tau \mathbf{S}_1\times\mathbf{S}_2, \sigma^2\mathbf{S}_2 + \tau^2\mathbf{S}_1 + \tau \mathbf{S}_2\times\mathbf{S}_1 \Big), \label{two-body-map-app} \\
&\sigma^2 = \frac{1}{2} \Big(1 + \mathbf{S}_1\cdot \mathbf{S}_2 \Big), \quad \tau \in \mathbb{R}. \nonumber
\end{align}
The propagator is a non-linear invertible mapping on a cartesian product of two 2-spheres, that takes a pair of unit vectors and transforms them according to Eq.~(\ref{two-body-map-app}).\\

\noindent Eq.~(\ref{RLL_app}) can be proven by a direct calculation, taking into account the structure of Pauli matrices. It expands as:
\begin{align}
\label{RLL_expand}
&\Big(1 - \frac{1}{4} \frac{1}{\lambda \mu} \mathbf{S}_{2} \cdot \mathbf{S}_1 \Big)\one +  \Big( \frac{1}{2i \lambda} \mathbf{S}_{2} + \frac{1}{2i \mu} \mathbf{S}_{1} - \frac{i}{4 \lambda \mu} \mathbf{S}_{2} \times \mathbf{S}_1   \Big)  \cdot \pmb{\sigma} =  \nonumber \\
&\Big(1 - \frac{1}{4} \frac{1}{\lambda \mu} \mathbf{S}'_{2} \cdot \mathbf{S}'_1 \Big)\one +  \Big( \frac{1}{2i \mu} \mathbf{S}'_{2} + \frac{1}{2i \lambda} \mathbf{S}'_{1} - \frac{i}{4 \lambda \mu} \mathbf{S}'_{2} \times \mathbf{S}'_1   \Big) \cdot \pmb{\sigma},
\end{align}
where the following identity has been used,
$
(\vec{S}_1 \cdot \pmb{\sigma})( \vec{S}_2 \cdot \pmb{\sigma}) =  (\vec{S}_1 \cdot \vec{S}_2)\one +  i(\vec{S}_1 \times \vec{S}_2) \cdot \pmb{\sigma}
$.
Three Pauli matrices and the identity matrix form a basis of the space of $2 \times 2$ matrices. Consequently the terms proportional to the idenitity and the vector of Pauli matrices in Eq.~(\ref{RLL_expand}) must coincide independently, 
for the equation to hold. For the term proportional to the identity we have:
\begin{align}
&\tau := \lambda - \mu,  \nonumber\\
&\mathbf{S}'_{1} \cdot \mathbf{S}'_2 = \nonumber \\
& = \frac{1}{(\sigma^2 + \tau^2)^2} \Big(\sigma^2\mathbf{S}_1 + \tau^2\mathbf{S}_{2} + \tau \mathbf{S}_1\times\mathbf{S}_{2} \Big) \cdot \Big(\sigma^2\mathbf{S}_{2} + \tau^2\mathbf{S}_1 + \tau \mathbf{S}_{2} \times\mathbf{S}_1 \Big) = \nonumber \\
& = \frac{1}{(\sigma^2 + \tau^2)^2} \Big((\sigma^4 + \tau^4) \mathbf{S}_{1} \cdot \mathbf{S}_2  + 2\sigma^2 \tau^2 - \tau^2 ( \mathbf{S}_{2} \times  \mathbf{S}_{1})^2\Big) = \nonumber \\
& = \frac{1}{(\sigma^2 + \tau^2)^2} \Big((\sigma^4 + \tau^4) \mathbf{S}_{1} \cdot \mathbf{S}_2  + 2\sigma^2 \tau^2 \mathbf{S}_{1} \cdot \mathbf{S}_{2} \Big) = \nonumber \\
& = \frac{\mathbf{S}_{1} \cdot \mathbf{S}_2 }{(\sigma^2 + \tau^2)^2} \Big(s\sigma^4 + \tau^4  + 2\sigma^2 \tau^2 \Big) = \nonumber \\
&= \mathbf{S}_{1} \cdot \mathbf{S}_2,
\end{align}
from which it follows that the identitiy term of Eq.~(\ref{RLL_expand}) holds. It remains to deal with the term proportional to the vector of Pauli matrices. To this end we quickly extract the following relations from the definition of the two-body propagator:
\begin{align}
\tau &:= \lambda - \mu, \nonumber\\
\mathbf{S}'_{1} + \mathbf{S}'_{2} &= \mathbf{S}_{1} + \mathbf{S}_{2} , \nonumber\\
\mathbf{S}'_{1} - \mathbf{S}'_{2} &= \frac{\sigma^2 - \tau^2}{\sigma^2 + \tau^2} \Big(\mathbf{S}_{1} - \mathbf{S}_{2}\Big) + \frac{2\tau}{\sigma^2+ \tau^2} \mathbf{S}_{1} \times \mathbf{S}_{2},\nonumber\\
\label{uber_eqs}
\mathbf{S}'_{1} \times \mathbf{S}'_{2} &= \frac{-2 \tau \sigma^2}{\sigma^2 + \tau^2}  \Big(\mathbf{S}_{1} - \mathbf{S}_{2}\Big)  + \frac{\sigma^2 - \tau^2}{\sigma^2 + \tau^2} \mathbf{S}_1 \times \mathbf{S}_{2},
\end{align}
where we have used a  formula for the triple vector product:
$\mathbf{a} \times (\mathbf{b} \times \mathbf{c}) = \mathbf{b} (\mathbf{a} \cdot \mathbf{c}) - \mathbf{c} (\mathbf{a} \cdot \mathbf{b})$.
The above equations show that the sum of spins is conserved whereas their difference and vector produt are interconnected. Bearing these relations in mind we rewrite the last term of Eq.~(\ref{RLL_expand}) as:
\begin{align}
&\Big( \frac{1}{2i \mu} \mathbf{S}'_{2} + \frac{1}{2i \lambda} \mathbf{S}'_{1} - \frac{i}{4 \lambda \mu} \mathbf{S}'_{2} \times \mathbf{S}'_1   \Big) = \nonumber \\
&= \Big(\frac{1}{4i}(\frac{1}{\lambda} + \frac{1}{\mu}) (\mathbf{S}'_{1} + \mathbf{S}'_{2}) + \frac{1}{4i}(\frac{1}{\lambda} - \frac{1}{\mu}) (\mathbf{S}'_{1} - \mathbf{S}'_{2})  + \frac{i}{4\lambda \mu}  \mathbf{S}'_{1} \times \mathbf{S}'_{2}  \Big).
\end{align}
Using Eqs.~(\ref{uber_eqs}) we easily rewrite the last expression using the initial spins:
\begin{equation}
 \Big(\frac{1}{4i}(\frac{1}{\lambda} + \frac{1}{\mu}) (\mathbf{S}_{1} + \mathbf{S}_{2}) + \frac{1}{4i}(\frac{1}{\mu} - \frac{1}{\lambda}) (\mathbf{S}_{1} - \mathbf{S}_{2})  + \frac{i}{4\lambda \mu}  \mathbf{S}_{1} \times \mathbf{S}_{2}  \Big).
\end{equation}
This is precisely the coefficient of the Pauli vector term of Eq.~(\ref{RLL_expand}) and completes the proof of the RLL equation.

\section{appendix: Self-duality}
\label{self-dual_append}
The elementary temporal dynamics of the model is given by:
\begin{align}
\label{selfdual_eq0}
\big(\mathbf{S}'_1, \mathbf{S}'_2 \big) &= \frac{1}{\sigma^2 + \tau^2}\Big(\sigma^2 \mathbf{S}_1 +  \tau^2 \mathbf{S}_2 + \tau \mathbf{S}_1 \times \mathbf{S}_2,\tau^2 \mathbf{S}_1 +  \sigma^2 \mathbf{S}_2 + 
\tau \mathbf{S}_2 \times \mathbf{S}_1\Big), \\
\sigma^2 &= \frac{1}{2}\big(1 +  \mathbf{S}_1\cdot\mathbf{S}_2\big), \quad \tau \in \mathbb{R}. \nonumber
\end{align}
In this appendix we answer the following question: Given the above temporal dynamics is it possible to derive the corresponding spatial dynamics? That is, is it possible to compute the pair  $\big(\mathbf{S}'_2, \mathbf{S}_2 \big)$ using a known pair of spins $\big(\mathbf{S}'_1, \mathbf{S}_1 \big)$? We show that this is indeed possible and that such a map is unique.  

If the vectors $\mathbf{S}_1, \mathbf{S}'_1$ are linearly dependent, the dynamics are trivial. When the pair is linearly independent, the vectors   $\mathbf{S}_1, \mathbf{S}'_1,  \mathbf{S}_1 \times \mathbf{S}'_1$ span the space.
We can therefore expand the vector $\mathbf{S}_2$ in this (in general) non-orthogonal basis:
\begin{equation}
\label{selfdual_ansatz}
\mathbf{S}_2 = A_1\mathbf{S}_1 + A_2\mathbf{S}'_1 + A_3\mathbf{S}_1 \times \mathbf{S}'_1,
\end{equation}
where $A_j$ are as of yet unknown real coefficients. 
From the known pair of vectors $\mathbf{S}_1, \mathbf{S}'_1$ and the definition of the two-body propagator (\ref{selfdual_eq0}) we can compute their scalar product $r = \mathbf{S}_1 \cdot \mathbf{S}'_1$ and calculate:
\begin{equation}
\mathbf{S}_1 \cdot \mathbf{S}_2 = \frac{(1+2\tau^2)r - 1}{(1 + 2\tau^2)-r}.
\end{equation}
Using the above relation a direct calulation also gives the following scalar prodcuts $\mathbf{S}_2 \cdot \mathbf{S}'_1$,  $\mathbf{S}_2 \cdot (\mathbf{S}_1 \times \mathbf{S}'_1)$:
\begin{align}
\mathbf{S}_2 \cdot \mathbf{S}_1 &= C_1 = \frac{(1+2\tau^2)r - 1}{(1 + 2\tau^2)-r},\nonumber\\
\mathbf{S}_2 \cdot \mathbf{S}'_1 &= C_2 = \frac{r^2 - r + 2\tau^2}{(1 + 2\tau^2)-r}, \nonumber\\
\mathbf{S}_2 \cdot (\mathbf{S}_1 \times \mathbf{S}'_1) &= C_3 = \frac{-2\tau(1-r^2)}{(1 + 2\tau^2)-r}. \label{selfdual_system}
\end{align}
Inserting the ansatz  (\ref{selfdual_ansatz}) for $\mathbf{S}_2$ gives us a connection between the coefficients  $A_j$ and $C_j$. Since the vector $\mathbf{S}_1 \times \mathbf{S}'_1$ is orthogonal to the vectors $\mathbf{S}_1$ and  $\mathbf{S}'_1$ coefficients $A_3$ and $C_3$ do not intermingle with other coefficients. 
The first two equations of the system (\ref{selfdual_system}) specify the mixing between coeffcieints $A_1, A_2, C_1, C_2$, which can be cast as a system of equations:
\begin{equation}
 \begin{bmatrix} C_1  \\ C_2  \end{bmatrix}
 =
  \begin{bmatrix}
   1 &
   r\\
    r &
   1 
   \end{bmatrix}
  \begin{bmatrix} A_1  \\ A_2 \end{bmatrix},
\end{equation}
with the solution:
\begin{equation}
 \begin{bmatrix} A_1  \\ A_2  \end{bmatrix}
 =
\frac{1}{1-r^2}
  \begin{bmatrix}
   1 &
   -r\\
    -r &
   1 
   \end{bmatrix}
  \begin{bmatrix} C_1  \\ C_2 \end{bmatrix}.
\end{equation}
The third equation of the system (\ref{selfdual_system}) gives:
\begin{equation}
A_3 =  \frac{C_3}{1 - r^2}.
\end{equation}
Knowing the values of coefficients $A_i$, we insert them into the ansatz (\ref{selfdual_ansatz}) and after some simple calculations arrive at:
\begin{equation}
\mathbf{S}_2 = \frac{1}{\delta^2 + \tau^2}\Big(-\delta^2 \mathbf{S}_1 +  \tau^2 \mathbf{S}'_1 - \tau \mathbf{S}_1 \times \mathbf{S}'_1\Big), \quad \delta^2 = \frac{1}{2}\big(1 -  \mathbf{S}_1\cdot\mathbf{S}'_1\big).
\end{equation}
Since the map (\ref{two-body-map}) preserves the sum of the pair of vectors:
\begin{equation}
\mathbf{S}_1 + \mathbf{S}_2 = \mathbf{S}'_1 + \mathbf{S}'_2,
\end{equation}
we can also easily compute the vector $\mathbf{S}'_2$. Taken together we have:
\begin{align}
\label{selfdual_eq1}
\big(\mathbf{S}_2 ,\mathbf{S}'_2 \big) &= \frac{1}{\delta^2 + \tau^2}\Big(-\delta^2 \mathbf{S}_1  + \tau^2 \mathbf{S}'_1 - \tau \mathbf{S}_1 \times \mathbf{S}'_1, 
-\delta^2 \mathbf{S}'_1 +  \tau^2 \mathbf{S}_1 - \tau \mathbf{S}_1 \times \mathbf{S}'_1\Big), \\
\delta^2 &= \frac{1}{2}\big(1 -  \mathbf{S}_1\cdot\mathbf{S}'_1\big).\nonumber
\end{align}
A simple calculation shows that such a map preserves the unit norm of the pair of vectors. Eq.~(\ref{selfdual_eq1}) specifies a unique spatial propagator, dual to the temporal propagator (\ref{selfdual_eq0}). \\
Moreover, flipping the signs of vectors $\mathbf{S}_1, \mathbf{S}'_2$ in Eq.~(\ref{selfdual_eq1}):
\begin{equation}
\tilde{\mathbf{S}}_1 = -\mathbf{S}_1, \quad \tilde{\mathbf{S}}'_2 = -\mathbf{S}'_2,
\end{equation}
the equation (\ref{selfdual_eq1}) is transformed into:
\begin{align}
\label{selfdual_eq2}
\big(\tilde{\mathbf{S}}_2,\mathbf{S}'_2\big) &= \frac{1}{\sigma^2 + \tau^2}\Big( \sigma^2 \mathbf{S}_1+ \tau^2 \tilde{\mathbf{S}}'_1 + \tau \mathbf{S}_1\times \tilde{\mathbf{S}}'_1, \sigma^2 \tilde{\mathbf{S}}'_1 + \tau^2 \mathbf{S}_1 + \tau \tilde{\mathbf{S}}'_1 \times \mathbf{S}_1\Big), \\
\nonumber
\sigma^2 &= \frac{1}{2}\big(1 +  \tilde{\mathbf{S}}_1\cdot\mathbf{S}'_1\big), \quad \tau \in \mathbb{R}.
\end{align}
Since this map is of exactly the same form as the temporal map (\ref{selfdual_eq0}), the map is said to be \textit{space-time self-dual} (see Fig.~\ref{fig:FigSTDual} of the main text). 
We can likewise show that flipping the sign of the other pair of vectors $\mathbf{S}'_1, \mathbf{S}_2$ in Eq.~(\ref{selfdual_eq1}) transforms it into:
\begin{align}
\label{selfdual_eq3}
\big(\tilde{\mathbf{S}}_2,\mathbf{S}'_2 \big) &= \frac{1}{\sigma^2 + \tau^2}\Big(\sigma^2 \mathbf{S}_1 + \tau^2 \tilde{\mathbf{S}}'_1 - \tau  \mathbf{S}_1\times\tilde{\mathbf{S}}'_1, 
\sigma^2 \tilde{\mathbf{S}}'_1 +  \tau^2 \mathbf{S}_1 - \tau \tilde{\mathbf{S}}'_1 \times \mathbf{S}_1\Big), \\
\nonumber
\sigma^2 &= \frac{1}{2}\big(1 +  \mathbf{S}_1\cdot \tilde{\mathbf{S}}'_1\big), \quad \tau \in \mathbb{R}.
\end{align}
This map once again has the same form as the original map (\ref{selfdual_eq0}), except that the value of the time parameter has been replaced by its negative, $\tau \rightarrow - \tau$.

\end{document}